\documentclass[a4paper,fleqn]{cas-dc}

\usepackage[numbers]{natbib}

\usepackage{amsmath,amsfonts}
\usepackage{algorithm}
\usepackage{array}
\usepackage[caption=false]{subfig}
\usepackage{textcomp}
\usepackage{stfloats}
\usepackage{url}
\usepackage{verbatim}
\usepackage{graphicx}

\usepackage[noend]{algpseudocode}
\usepackage[section]{placeins}

\usepackage{orcidlink}

\usepackage{booktabs}     
\usepackage{tabu,adjustbox}
\usepackage{multirow,rotating}

\usepackage{csquotes}

\begin{document}
\let\WriteBookmarks\relax
\def\floatpagepagefraction{1}
\def\textpagefraction{.001}
\shorttitle{I2I-PR: Deep Iterative Refinement for Phase Retrieval using Image-to-Image Diffusion Models}
\shortauthors{M.O. Kaya and F.S. Oktem}

\title [mode = title]{I2I-PR: Deep Iterative Refinement for Phase Retrieval using Image-to-Image Diffusion Models}



\author[1,2]{Mehmet Onurcan Kaya}[type=editor,
                        orcid=0009-0006-2606-3992
                        ]
\cormark[1]
\ead{monka@dtu.dk}

\credit{Conceptualization, Methodology, Software, Writing – original draft}


\affiliation[1]{organization={Department of Electrical Engineering, Middle East Technical University (METU)},
                city={Ankara},
                postcode=06800, 
                country={Turkey}}
                
\affiliation[2]{organization={Department of Applied Mathematics and Computer Science, Technical University of Denmark},
                city={Kongens Lyngby},
                postcode=2800, 
                country={Denmark}}


\author[1]{Figen S. Oktem}[%
   orcid=0000-0002-7882-5120
   ]
\ead{figeno@metu.edu.tr}

\credit{Conceptualization, Supervision, Writing – review and editing}



\cortext[cor1]{Corresponding author}


\begin{abstract}
Phase retrieval aims to recover a signal from intensity-only measurements, a fundamental problem in many fields such as imaging, holography, optical computing, crystallography, and microscopy.
Although there are several well-known phase retrieval algorithms, including classical alternating projection-based solvers, the reconstruction performance often remains sensitive to initialization and measurement noise. Recently, 
diffusion models have gained traction in various image reconstruction tasks, yielding significant theoretical insights and practical advances. 
In this work, we introduce a deep iterative refinement framework that redefines the role of diffusion models in phase retrieval. 
Instead of generating images from random noise, our method starts with multiple physically consistent initial estimates and iteratively refines them through a learned image-to-image diffusion process. This enables data-driven phase retrieval that is both interpretable and robust, leveraging the strengths of classical solvers while mitigating their weaknesses.
Furthermore, we propose 
an enhanced initialization strategy that integrates classical algorithms with a novel acceleration mechanism to obtain reliable initial estimates. During inference, we adopt a geometric self-ensemble strategy based on input flipping, together with output aggregation  
to further improve the final reconstruction quality. 
Comprehensive experiments demonstrate that our approach achieves substantial gains in both training efficiency and reconstruction quality, consistently outperforming classical and recent state-of-the-art methods. These results highlight the potential of diffusion-driven refinement as an effective and general framework for robust phase retrieval across diverse applications.
The source code and trained models are available at \href{https://github.com/METU-SPACE-Lab/I2I-PR-for-Phase-Retrieval}{https://github.com/METU-SPACE-Lab/I2I-PR-for-Phase-Retrieval}.
\end{abstract}




\maketitle

\section{Introduction}

Phase retrieval (PR) is a fundamental inverse problem in many scientific and engineering fields, where the goal is to recover a signal or image from only the intensity measurements, typically from its Fourier transform. This problem is particularly important in applications such as holography, microscopy, crystallography, and coherent diffractive imaging 
~\cite{shechtman2015phase,Dong_2023,personaFienup:13,1963_walther_pr_optics}.
The challenge in phase retrieval arises due to the loss of phase information, which carries the majority of structural information in an image \cite{oppenheim2005importance}. Consequently, the PR problem is non-linear and ill-posed. As a result, classical methods such as the Hybrid Input-Output (HIO) and Error Reduction (ER) algorithms rely heavily on initializations and specific constraints to converge to a solution. Although these methods can often produce good reconstructions, they are highly sensitive to noise and initialization, leading to suboptimal solutions, especially in 
noisy scenarios ~\cite{shechtman2015phase, marchesini2007invited}. Addressing this sensitivity is key to improving the reliability and performance of phase retrieval techniques.

In response to these challenges, deep learning has recently become a powerful tool for solving inverse problems in imaging, including PR \cite{lopeztaipa, Wang_2024}. These data-driven approaches, particularly those utilizing deep neural networks, have shown remarkable success in directly reconstructing images from measurements or refining initial estimates from classical methods. 
Despite their promise, many deep learning-based PR solutions face significant challenges, including their lack of robustness to noise as well as domain shifts between training and test data, which can result in poor generalization. Also, these models often suffer from a lack of interpretability and fail to provide uncertainty quantification, making it difficult to understand the reliability of the reconstructions. Further limitations include the need for extensive parameter tuning and sensitivity to hyperparameters \cite{lopeztaipa, Wang_2024,isil2024deep}, which complicates their deployment in real-world applications. These limitations highlight the need for more efficient, adaptable models.

Numerous schemes have emerged to harness deep learning for solving inverse problems, with unrolling methods being particularly relevant to our work. Unrolling methods aim to replicate traditional iterative algorithms through a series of trainable network layers. However, they often suffer from prolonged training times and computational inefficiencies \cite{Wang_2024}. Additionally, many existing unrolling phase retrieval methods initiate the reconstruction from random noise, resulting in suboptimal use of the denoiser's capacity and slower convergence in training.
Classical diffusion pipelines also suffer from similar disadvantages since they initiate reconstruction from pure noise instead of using a warm-start with an initial crude estimate, underutilizing the denoiser's capacity and leading to slower convergence \cite{ddrmprKaya:25, 10477083, Bansal2022ColdDI, delbracio2023inversion}.
These inefficiencies limit their practical use, as extended training and computation times are not feasible for many real-world applications \cite{Wang_2024}. 
Addressing these inefficiencies requires innovative strategies to improve both the initialization process and the models themselves.

In contrast, recent image-to-image (I2I) diffusion models introduce a warm-start approach by initializing the reconstruction with a plausible estimate rather than pure noise. This refinement-based process facilitates learning by focusing on improving an initial estimate rather than generating an image from scratch. The effectiveness of I2I pipelines has been demonstrated in various applications, including deblurring and super-resolution~\cite{delbracio2023inversion, Saharia2021PaletteID, Bansal2022ColdDI, Whang2021DeblurringVS, liu2023flow, heitz2023iterative}.

In this work, we propose image-to-image-PR (I2I-PR), a data-driven phase retrieval approach that unrolls classical solvers into a deep iterative refinement framework inspired by modern I2I pipelines.
More specifically, our method extends the Inversion by Direct Iteration
(InDI)~\cite{delbracio2023inversion} framework to phase retrieval by introducing novel strategies for denoising, initialization, and measurement consistency. 
Unlike conventional diffusion-based methods that begin with random initial guesses, our approach starts from 
multiple initial estimates
provided by classical solvers and iteratively refines them using the learned I2I model. This hybridization significantly improves both the robustness and quality of the reconstruction, especially in the presence of noise and difficult measurement conditions.
During inference, we adopt a geometric self-ensemble strategy based on input flipping, together with output aggregation to further improve reconstruction quality. This deep iterative framework combines the strengths of classical solvers with the learned priors from diffusion models, enhancing both training time and reconstruction performance.

Our methodology marks a significant departure from conventional approaches by leveraging advanced denoising strategies combined with novel initialization and aggregation techniques. Similar to unrolling methods, instead of treating the problem as a single-step inversion, our model iteratively improves the estimate at each step by effectively decomposing the main phase retrieval problem into smaller subproblems that exhibit reduced ill-posedness.
Moreover, by initiating the recovery process from a plausible estimate rather than random noise, our approach more efficiently utilizes the denoiser's model capacity and significantly reduces training time compared to conventional unrolling methods.

The primary contributions of our work can be summarized as follows:
\begin{itemize}
    \item Diverging from other diffusion-based PR methods, our learned image-to-image framework starts with multiple plausible initial estimates and refines them, thereby more effectively utilizing the denoiser's capacity and reducing overall training duration.
    \item We have integrated a novel accelerated error reduction algorithm into our initialization strategy, significantly enhancing the robustness and speed of convergence for the phase retrieval process.
    \item Our aggregation technique combines multiple reconstructions during inference to improve distortion metrics while also substantially enhancing the perceptual quality of the reconstructed images. 
\end{itemize}

The integration of I2I denoising pipelines with model-based approaches holds promise for developing robust and reliable stochastic solvers for phase retrieval as well as for nonlinear inverse problems more broadly. Comprehensive experiments demonstrate that our method provides superior performance compared to both classical and recent techniques. The results illustrate its effectiveness in addressing the inherent challenges of phase retrieval. Some preliminary results of
this research were previously presented in \cite{kaya2025_eusipco}.

The structure of this paper is organized as follows: Section \ref{sec:prproblemdefinition} provides the mathematical formulation of the phase retrieval problem. Section \ref{sec:relatedworksindi} reviews existing research that has shaped the development of our approach. Our developed approach, including the novel aspects of our framework, is presented in Section \ref{sec:methodindi}. Comparative analyses of our method against both classical techniques and recent deep learning-based approaches are presented in Section \ref{sec:resultsindi}. Finally, Section \ref{sec:conclusionindi} summarizes our key findings and outlines potential directions for future research in this dynamic field.

\section{Phase Retrieval Problem}
\label{sec:prproblemdefinition}
In its broadest sense, the phase retrieval problem involves reconstructing an unknown signal \(\mathbf{x}\) from the intensity measurements expressed as
\begin{equation}
\mathbf{y^2} = \mathbf{\vert Ax \vert^2 + w}
\label{eq:eq1general}
\end{equation}
where $\mathbf{y^2} \in \mathbb{R}^{m}$ denotes the noisy intensity measurements in vector form, \(\mathbf{A} \in \mathbb{C}^{m \times n}\) represents a known linear operator specific to the application, and $\mathbf{x} \in \mathbb{C}^{n}$ denotes the vectorized form of the $\sqrt{n} \times \sqrt{n}$ unknown image.
Moreover, $|\cdot|^2$ denotes element-wise magnitude squaring operation applied to a complex-valued vector.
The term $\mathbf{w} \in \mathbb{R}^{m}$ denotes the measurement noise which is generally modeled as Poisson-distributed, but a normal approximation can be used in many cases~\cite{pmlr-v80-metzler18a}. Formally, $\mathbf{w} \sim \mathcal{N}(\mathbf{0}, \alpha^2 \text{diag}(\vert \mathbf{Ax} \vert^2))$ where $\alpha$ is a scaling factor that adjusts the signal-to-noise ratio (SNR).

A key special case is the classical Fourier phase retrieval problem, with $\mathbf{A} = \mathbf{F}$ being Fourier matrix and the unknown image $\mathbf{x}$ assumed to be real-valued, non-negative, and of finite support.

It is worth mentioning that phase retrieval problems are more difficult to solve
than linear inverse problems due to the non-linearity in this forward model. This problem has many important applications in imaging~\cite{1963_walther_pr_optics}, computer-generated holography~\cite{latychevskaia2019iterative},
optical computing~\cite{chang2018hybrid}, crystallography~\cite{millane1990phase}, microscopy~\cite{zheng2013wide}, speech processing~\cite{peer2022beyond}, optical engineering~\cite{hubbleFienup:93}, and theoretical machine learning~\cite{sarao2020optimization}, to name a few. Despite their diverse applications and different physical setups, the forward models in phase retrieval converge to a common mathematical formulation as given in Eq.~\ref{eq:eq1general}
~\cite{shechtman2015phase,Dong_2023,personaFienup:13}.

\section{Related Work}
\label{sec:relatedworksindi}

\subsection{Classical Iterative Projection Techniques} 

\label{sec:Alter}

Iterative projection techniques are well-known and fundamental tools for phase retrieval. One of the earliest and most well-known algorithms is the classical Gerchberg-Saxton (GS) algorithm~\cite{gs1978}, which iteratively applies magnitude constraints in both the spatial and measurement domains to reconstruct an unknown signal. An enhancement of the GS algorithm is the Error Reduction (ER) algorithm, which incorporates additional spatial domain constraints beyond just magnitude~\cite{fienup1978reconstruction}. The ER iterations are mathematically expressed as follows:
\begin{equation}\label{eq:prnetttttttthio2}
\begin{aligned}
\mathbf{x}_{k+1}[n] = \left\{ \begin{array}{rcl}
\mathbf{x}_k'[n] & \text{for} & n \notin \gamma \\
0 & \text{for} & n \in \gamma \\
\end{array}\right.
\end{aligned}
\end{equation}
where
\begin{equation}\label{eq:prnettttttthio1}
\begin{aligned}
\mathbf{x}_k' = \mathbf{A}^{\dagger}\left\{\mathbf{y} \odot \frac{\mathbf{A} \mathbf{x}_k}{\vert \mathbf{A} \mathbf{x}_k \vert}\right\}.
\end{aligned}
\end{equation}

In these equations, $\mathbf{x}_k \in \mathbb{R}^{m}$ represents the reconstruction at the $k^{\text{th}}$ iteration, $\mathbf{A^{\dagger}}$ denotes the pseudoinverse of the forward matrix, $\odot$ signifies element-wise multiplication, and $\gamma$ is the set of indices $n$ where $\mathbf{x}_k'[n]$ fails to meet the spatial domain constraints~\cite{fienup1982comparison}.

A particularly significant and widely used method among alternating projection techniques is the Hybrid Input-Output (HIO) algorithm~\cite{fienup1982comparison}, which builds upon the principles of the ER algorithm.
In the HIO method, magnitude constraints and various spatial domain constraints (such as support, non-negativity, and real-valuedness) are iteratively applied, similar to the ER algorithm. However, the key distinction is that HIO does not force the iterates to strictly satisfy the constraints at every step. Instead, it uses the iterates to progressively guide the algorithm towards a solution that meets the constraints~\cite{fienup1982comparison}. The HIO iterations are mathematically expressed as follows:
\begin{equation}\label{eq:prnetttthio2}
\begin{aligned}
\mathbf{x}_{k+1}[n] = \left\{ \begin{array}{rcl}
\mathbf{x}_k'[n] & \text{for} & n \notin \gamma \\
\mathbf{x}_{k}[n] - \beta \mathbf{x}_k'[n] & \text{for} & n \in \gamma \\
\end{array}\right.
\end{aligned}
\end{equation}
where
\begin{equation}\label{eq:prnetttthio1}
\begin{aligned}
\mathbf{x}_k' = \mathbf{A}^{\dagger}\left\{\mathbf{y} \odot \frac{\mathbf{A} \mathbf{x}_k}{\vert \mathbf{A} \mathbf{x}_k \vert}\right\}.
\end{aligned}
\end{equation}

Here, $\beta$ is a constant parameter (commonly set to 0.9)~\cite{fienup1982comparison}.

Despite the lack of a comprehensive theoretical understanding of the HIO method's convergence behavior, it has been empirically observed to converge to acceptable solutions in a wide array of applications. However, the reconstructions produced by HIO can sometimes contain artifacts and errors. These issues are often attributed to the algorithm getting trapped in local minima or to the amplification of noise within the solution~\cite{shechtman2015phase, marchesini2007invited}. To address these limitations, numerous variations and enhancements of the HIO method have been proposed, aiming to improve its reconstruction performance and reliability~\cite{stefanoqianpty, Maiden:17}.

The geometric interpretation of the Error Reduction (ER) and Hybrid Input-Output (HIO) algorithms provides a clear visualization of their operational mechanics in the context of phase retrieval. Both algorithms utilize iterative projections between constraint sets defined in different domains, but they differ significantly in their approach to managing deviations from these constraints.

The ER algorithm applies consecutive projection operations to obtain the estimate iteratively, aligning it within the intersection of spatial and magnitude constraints. Mathematically, this is represented by:
\begin{equation}
\mathbf{x}_{k+1} = \mathcal{P}_S \mathcal{P}_F \mathbf{x}_{k}
\end{equation}
where \(\mathcal{P}_F\) and \(\mathcal{P}_S\) are projection operators enforcing magnitude and space domain constraints, respectively. Geometrically, if all constraints were convex, this sequence of projections would direct the estimate towards the intersection of the constraint sets in a monotonic, stepwise manner.

Differently, HIO incorporates a reflective step to handle violations of constraints, allowing for correction of the trajectory during iterations. The iteration formula for HIO is expressed as:
\begin{equation}
\mathbf{x}_{k+1} = \left[\frac{\mathbf{I} + \mathcal{R}_S \mathcal{R}_F}{2} + (1-\beta)(\mathbf{I}-\mathcal{P}_S)\mathcal{P}_F\right] \mathbf{x}_{k}
\end{equation}
where \(\mathcal{R}_S = 2\mathcal{P}_S - \mathbf{I}\) and \(\mathcal{R}_F = 2\mathcal{P}_F - \mathbf{I}\) denote reflection operators related to the space and measurement domain constraints, respectively \cite{Fannjiang2020TheNO}. This formulation introduces a dynamic adjustment to the trajectory, allowing the method to navigate around potential local minima and avoid stagnation, a common limitation of simpler projection methods in nonconvex feasibility problems.

The ER algorithm follows a direct approach toward the solution by closely following the constraints, while HIO allows for a more explorative strategy, potentially circumventing issues such as local minima through its reflective and corrective steps. These interpretations explain the distinct types of pathways each algorithm takes in the constrained solution landscape of phase retrieval.

Notably, when \(\beta=1\), HIO becomes equivalent to the Douglas-Rachford algorithm \cite{AragnArtacho2019TheDA}. This equivalence is particularly significant as the Douglas-Rachford algorithm is known for its efficacy in handling nonconvex feasibility problems, such as phase retrieval. However, while HIO offers advantages in avoiding local minima due to its more explorative update strategy, it can exhibit challenges such as spiraling dynamics during convergence \cite{AragnArtacho2019TheDA}.

\subsection{Image-to-Image Pipelines for Inverse Problems}

Many data-driven techniques for solving inverse problems, such as unrolled methods, are often criticized for their computational and memory inefficiencies as well as slow training, primarily due to their extensive computational requirements and the iterative refinement they employ \cite{heckel2024deep}. Also, classical diffusion pipelines for solving inverse problems typically initiate the reconstruction/restoration from a random noise image~\cite{ddrmprKaya:25, 10477083}. This starting point is generally not ideal because it does not exploit a crude reconstruction that can be obtained with a simple computation. This could potentially facilitate faster convergence to reconstruction and avoid the waste of the denoiser's model capacity.

On the other hand, recent image-to-image pipelines initiate the reconstruction process with a warm start. Starting with a plausible estimate rather than a complete noise image facilitates training since it is only required to learn how to refine this initial estimate. 
This "image-to-image" refinement idea instead of "noise-to-image" pipeline is becoming prevalent in the literature \cite{delbracio2023inversion, Saharia2021PaletteID, Bansal2022ColdDI, Whang2021DeblurringVS, liu2023flow, heitz2023iterative}.

One such image-to-image pipeline is the Inversion by Direct Iteration (InDI) method which simplifies model training by defining a specific diffusion process from an initial estimate \cite{delbracio2023inversion}. 
In particular, in its stochastic version, the InDI method employs a sophisticated approach for image restoration problems by integrating denoising diffusion probabilistic models into its framework. This methodology leverages the incremental improvement of image quality through iterative refinement and denoising. Each refinement is modified by stochastic perturbations, which are essential for handling various degradation levels and ensuring robustness in the refinement and denoising process.

The InDI method provides an innovative approach to supervised image restoration that mitigates the "regression to the mean" effect, yielding images that are more realistic and detailed compared to traditional regression-based techniques. This method improves image quality incrementally in small steps, similar to generative denoising diffusion models \cite{kawar2022denoising}. Traditional single-step regression models often produce averaged outputs that lack detail and realism due to the ill-posed nature of the problem, where multiple high-quality images can plausibly reconstruct a given low-quality input \cite{Blau2017ThePT}. InDI’s strength lies in its iterative refinement process, which enhances perceptual quality by gradually improving the inferred image. Unlike generative denoising diffusion models that need prior knowledge of the degradation process, InDI learns the restoration directly from the paired examples of low-quality noisy images and high-quality clean images. This approach has been applied to various image degradation scenarios such as deblurring and super-resolution, making it a versatile and powerful solution for image restoration tasks \cite{delbracio2023inversion}.

The training of the denoising model within the InDI framework is strategically designed to cope with varied degradation and noise levels determined by the defined fixed degradation schedule. During training, this noise/degradation process is simulated with computationally cheap computations. The model employs the following equation for the intermediate degraded
image \cite{delbracio2023inversion}:
\begin{equation}
\mathbf{x}_t = (1 - t)\mathbf{x} + t\mathbf{z} + t\sigma_t\boldsymbol{\epsilon},
\end{equation}
where \(\mathbf{x}\) represents the clean target image, \(\mathbf{z}\) is the low-quality input, \(t\) ranges from 0 to 1, indicating the transition from clean to noisy image, \(\sigma_t\) varies with \(t\) as the standard deviation of noise, and lastly \(\boldsymbol{\epsilon}\), following a standard normal distribution \(\mathcal{N}(\mathbf{0}, \mathbf{I})\), introduces stochasticity.
During training, this fixed degradation schedule formulation is used to simulate $\widehat{\mathbf{x}}_t$, the output of the iterative refinement process at $t^{\text{th}}$ timestep similar to teacher forcing. After training, the image degraded according to the defined fixed degradation schedule ${\mathbf{x}}_t$ will be very close to $\widehat{\mathbf{x}}_t$ if the iterative refinement process is successfully learned due to the mathematical properties of this definition. Thus, there is no problem using the computationally cheap fixed degradation schedule to simulate $\widehat{\mathbf{x}}_t$.
This formulation prepares the model to reverse noise effects by optimizing the neural network parameters \(\boldsymbol{\theta}\) as follows \cite{delbracio2023inversion}:
\begin{equation}
\underset{\boldsymbol{\theta}}{\operatorname{minimize}} \mathbb{E}_{\mathbf{x}, \mathbf{z} \sim p(\mathbf{x}, \mathbf{z})} \left[ \mathbb{E}_{t \sim p(t)}\left[\|\text{Denoiser}_{\boldsymbol{\theta}}(\mathbf{x}_t, t) - \mathbf{x}\|_2^2\right]\right],
\end{equation}
which is the mean squared error between the denoised image and the original clean image across randomly sampled noise levels \(t\).

Once trained, the model uses the denoising function to iteratively restore the noisy image towards its original state. The guiding recurrence relation for this inference is:
\begin{equation}
\label{eq:indiinference}
\widehat{\mathbf{x}}_{t-\tau} = \left(1 - \frac{\tau}{t}\right) \widehat{\mathbf{x}}_t + \frac{\tau}{t} \operatorname{Denoiser}(\widehat{\mathbf{x}}_t, t) + (t-\tau) \sqrt{\sigma_{t-\tau}^2 - \sigma_t^2} \boldsymbol{\epsilon},
\end{equation}
where \(\tau\) is a small step back in time from \(t\), enhancing the restoration precision at each step. The term \(\sigma_{t-\tau}^2 - \sigma_t^2\) reflects the decrease in noise variance, aiding in the gradual refinement/restoration process. This process starts from $t = 1$ with $\widehat{\mathbf{x}}_{1} = \mathbf{z} + \sigma_1 \mathbf{w}$ where  $\mathbf{w} \sim \mathcal{N}(\mathbf{0},\mathbf{I})$. At each step, a new noise $\boldsymbol{\epsilon}$ is sampled and added to the current state.

The denoising function \(\operatorname{Denoiser}(\widehat{\mathbf{x}}_t, t)\) computes the expected clean image using the current noisy estimate $\widehat{\mathbf{x}}_t$, mirroring the conditional expectation of the posterior distribution \(\mathbb{E}[\mathbf{x}_{t-1} \mid \mathbf{x}_t]\). This function is optimized during training to minimize the reconstruction error and encapsulates the core denoising capability of the model.

The InDI framework, under certain conditions, is equivalent to other image-to-image pipelines such as Schrödinger Bridge \cite{Liu2023I2SBIS, chung2024direct} and Cold Diffusion \cite{Bansal2022ColdDI, delbracio2023inversion}. This equivalence reflects the versatile and robust nature of the InDI method since it aligns with other established methodologies that worked well for different reconstruction problems in imaging such as MRI reconstruction \cite{Mirza2023LearningFD}.

\section{Developed Method}
\label{sec:methodindi}


Our method builds upon the Inversion by Direct Iteration (InDI) framework, refining the initialization stage output to improve robustness and reconstruction quality in phase retrieval (PR). By leveraging a hybrid iterative initialization process that incorporates both the Hybrid Input-Output (HIO) and Error Reduction (ER) methods with a novel acceleration mechanism, our approach generates a crude initial estimate. This estimate is then iteratively refined using a learned image-to-image (I2I) diffusion pipeline based on InDI, improving the reconstruction through successive denoising and data consistency stages.

In this section, we provide a detailed explanation of our approach, starting with the enhanced initialization stage, followed by the iterative refinement using the diffusion model, and concluding with geometric self-ensemble and aggregation strategies to further improve reconstruction quality. This hybrid framework combines the strengths of classical solvers with modern I2I refinement techniques, enabling both efficient training and superior performance in challenging PR scenarios.

\subsection{Iterative Refinement Stage through Inversion by Direct Iteration}

\begin{figure*}[ht!]
\centering\includegraphics[width=\textwidth, page=1]{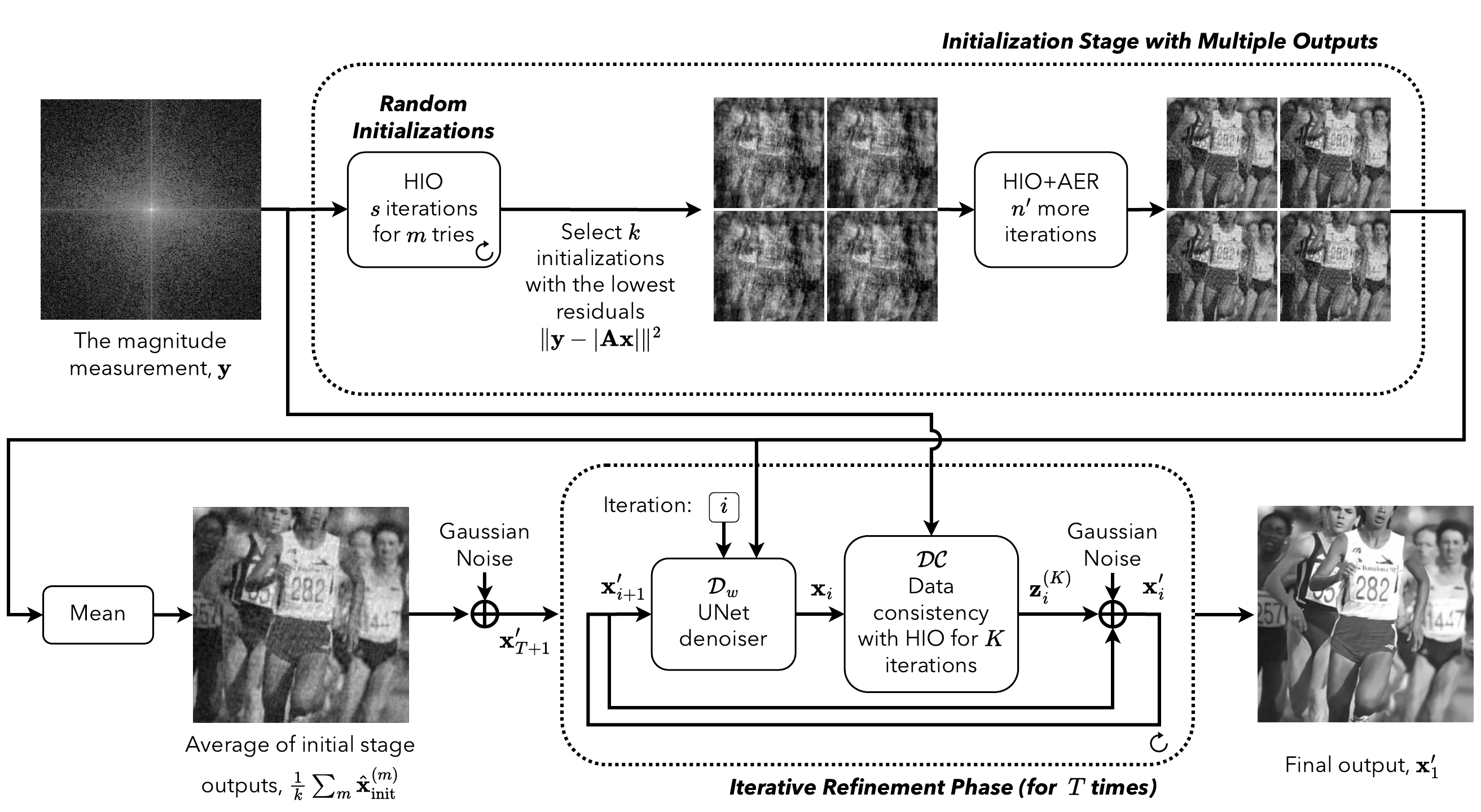}
\caption{The overall pipeline of I2I-PR. The method begins with an initialization stage combining HIO and AER to generate multiple candidate solutions. These are refined through an iterative image-to-image denoising process that incorporates data consistency with classical HIO updates.}
\label{fig:indinononooverallpipeline}
\end{figure*}

Our initialization procedure generates \(k\) distinct crude estimates \(\{ \hat{\mathbf{x}}_{\text{init}}^{\textit{(m)}} \}_{m = 1}^k\) for the same measurement \(\mathbf{y}\). This set of diverse estimates is essential for providing robustness and ensuring that our model has a broad foundation for refinement. As each crude estimate may correctly recover different regions while others may still contain artifacts, the model may leverage the strengths of each estimate. In our framework, which builds upon the InDI method, we initiate the iterative refinement process by computing the mean of these crude estimates. This mean serves as our starting point for refinement and is calculated as $
\mathbf{z} = \frac{1}{k} \sum_{m=1}^{k} \hat{\mathbf{x}}_{\text{init}}^{\textit{(m)}},
$
where \(\mathbf{z}\) is the initial estimate that aggregates the information from the \(k\) individual crude estimates.

In contrast to the single-image approach originally proposed in the InDI framework, where the iterative refinement begins from one initialization, our method generalizes this by incorporating multiple estimates. By taking the mean of the \(k\) crude estimates, we aim to provide a more stable and reliable initialization, reducing the risk of bias from any single noisy estimate. This aggregation effectively acts as a regularizer, smoothing out extreme variations and improving the overall quality of the starting point for the denoising process.

Following the formulation described in Eq. \ref{eq:indiinference}, we treat this initial mean \(\mathbf{z}\) as the noisy input \(\mathbf{z}\) in the InDI method. The target clean image \(\mathbf{x}\), as defined previously, remains the same, and the model is trained to progressively refine \(\mathbf{z}\) towards \(\mathbf{x}\) through a sequence of denoising steps. At each step, the model leverages both the noisy input \(\mathbf{z}\) and the timestep information to iteratively predict cleaner versions of the image, following the denoising process described earlier.

Averaging multiple initial reconstructions, while beneficial for providing a stable starting point, inevitably leads to some information loss, as fine details and distinct features present in individual estimates may get blurred. To mitigate this, we leverage a more refined approach: conditioning the denoiser on multiple initial reconstructions instead of solely relying on their average. This ensures that the model retains the unique contributions of each crude estimate, preserving important details that might otherwise be lost in the averaging process.
To counteract the information loss typically associated with this averaging, our empirical research has demonstrated that conditioning the denoiser on multiple initial reconstructions significantly improves reconstruction performance. Specifically, rather than using the standard input in the original InDI framework, which is $\operatorname{Denoiser}(\widehat{\mathbf{x}}_t, t)$, we incorporate a set of \(k\) initial reconstructions. Consequently, the denoiser now operates as $\operatorname{Denoiser}(\widehat{\mathbf{x}}_t, t, \{ \hat{\mathbf{x}}_{\text{init}}^{\textit{(m)}} \}_{m = 1}^k)$ at each step of the iterative refinement.
This configuration means that the denoiser takes 
$k+1$ inputs -- $k$ from the multiple initial reconstructions and one from the classical InDI current estimate $\widehat{\mathbf{x}}_t$ and produces one estimate. 
This approach utilizes additional information from the initial estimates, significantly enhancing the accuracy and efficacy of the image reconstruction and refinement process.

In image reconstruction tasks, balancing denoising and data consistency is crucial for producing high-quality results. In each iteration of our framework, we first perform a denoising step followed by a data consistency update using the HIO method.
Formally, rather than using the standard input in the original InDI framework, which is $\operatorname{Denoiser}(\widehat{\mathbf{x}}_t, t)$, we use $\operatorname{HIO}(\operatorname{Denoiser}(\widehat{\mathbf{x}}_t, t, \{ \hat{\mathbf{x}}_{\text{init}}^{\textit{(m)}} \}_{m = 1}^k), \mathbf{y})$.
Subsequently, Gaussian noise is added according to the InDI method to facilitate the iterative refinement process. Relying solely on denoising without incorporating physics-informed blocks can lead to outputs that are incompatible with the measurements, ultimately compromising reconstruction quality. While HIO provides significant advantages, it can also amplify noise and converge to local minima, resulting in artifacts in the final images. To address these challenges, our framework employs an iterative denoising-data consistency approach, similar to those used in many unrolling methods \cite{Aggarwal2017MoDLMD}. This strategy is designed to escape local minima and minimize artifacts, thereby enhancing the overall effectiveness of the reconstruction process. Note that for problems with no spatial constraints, HIO can be replaced by other alternating projection methods.

In summary, we enhance the original InDI framework presented in Eq. \ref{eq:indiinference} to improve its performance for phase retrieval. The modified framework is expressed as
\begin{equation}
\begin{aligned}
\label{eq:indiourversion}
\widehat{\mathbf{x}}_{t-\tau} = \left(1 - \frac{\tau}{t}\right) \widehat{\mathbf{x}}_t &+ \frac{\tau}{t} \operatorname{HIO}(\operatorname{Denoiser}(\widehat{\mathbf{x}}_t, t, \{ \hat{\mathbf{x}}_{\text{init}}^{\textit{(m)}} \}_{m = 1}^k), \mathbf{y}) \\
&+ (t-\tau) \sqrt{\sigma_{t-\tau}^2 - \sigma_t^2} \boldsymbol{\epsilon}
\end{aligned}
\end{equation}
Here, $\operatorname{HIO}(\operatorname{Denoiser}(\widehat{\mathbf{x}}_t, t, \{ \hat{\mathbf{x}}_{\text{init}}^{\textit{(m)}} \}_{m = 1}^k), \mathbf{y})$ represents the HIO operation, which utilizes a measurement (magnitude) derived from a convex combination of the original measurement $\mathbf{y}$ and the estimated measurement based on the denoised estimate at the current timestep. This measurement update strategy, inspired by prior work \cite{isil2024deep}, effectively combines the original measurement with the best available estimate to mitigate artifacts caused by measurement noise. Consequently, the proposed pipeline, illustrated in Fig. \ref{fig:indinononooverallpipeline}, achieves an efficient use of the denoiser's model capacity and strikes a balance between denoising and data consistency to produce high-quality results. By discretizing the timestep interval $[0, 1]$ into $T$ steps, we derive Algorithm \ref{alg:overallpipeline} from Eq. \ref{eq:indiourversion}.

\begin{algorithm}[tb]
\caption{I2I-PR}\label{alg:overallpipeline}
\textbf{Input:} $\mathbf{y}$ is the noisy magnitude measurements, $T, K, \beta, \{ \sigma_i \}_{i=0}^T$ are fixed hyperparameters, $\boldsymbol{\lambda} \in \mathbb{R}^T$ is a learnable vector, initialized with logarithmically increasing values


\begin{algorithmic}[1]
\Statex \textit{Initialization:}

\State $\{ \hat{\mathbf{x}}_{\text{init}}^{\textit{(m)}} \}_{m = 1}^k \gets$ Initialization procedure($\mathbf{y}$)

\State $\mathbf{w} \gets \text{sample from } \mathcal{N}(\mathbf{0},\mathbf{I}_n)$

\State $\mathbf{x}'_{T+1} \gets $$ \frac{1}{k} \sum_m \hat{\mathbf{x}}_{\text{init}}^{\textit{(m)}}$ + $\sigma_T \mathbf{w}$ 
\\


\Statex \textit{Main loop:}
\For{$i = T $ to $1$}
    \State $\mathbf{x}_i \gets $Denoiser($\mathbf{x}_{i+1}'$, $i$, $\{ \hat{\mathbf{x}}_{\text{init}}^{\textit{(m)}} \}_{m = 1}^k$)
    

    \State $\mathbf{z}_i^{(0)} \gets \mathbf{x}_i $

    \State $ {\mathbf{y}_i}' \gets \boldsymbol{\lambda}_i \mathbf{y} + (1-\boldsymbol{\lambda}_i) |\mathbf{Az}_i^{(0)}| $

	\For{$k=1$ to $K$}
    \State ${\mathbf{z}_i^{(k)}}' \gets \mathbf{A^{\dagger}} \left( {\mathbf{y}_i}' \odot \frac{\mathbf{Az}_i^{(k-1)}}{|\mathbf{Az}_i^{(k-1)}|} \right)  $ 

\State $\gamma \gets$ \parbox[t]{\dimexpr0.8\linewidth-\algorithmicindent}{the set of indices where ${\mathbf{z}_i^{(k)}}'$ violates space domain constraints (e.g., support and non-negativity)}
    
\State ${\mathbf{z}_i^{(k)}}[n] \gets 
\begin{cases} 
  {\mathbf{z}_{i}^{(k)}}'[n] & \text{, }  n \notin \gamma \\
  
    \mathbf{z}_{i}^{(k-1)}[n] - \beta {\mathbf{z}_{i}^{(k)}}'[n]
  
  & \text{, }  n \in \gamma
\end{cases}$

    \EndFor

\State $\boldsymbol{\epsilon} \gets \text{sample from } \mathcal{N}(\mathbf{0},\mathbf{I}_n)$

\State $\mathbf{x}_i' \gets \frac{1}{i} \mathbf{z}_i^{(K)}
+ \left(1 - \frac{1}{i}\right) \mathbf{x}_{i+1}'
+ \frac{i-1}{T} \sqrt{\sigma_{i-1}^2 - \sigma_i^2} \boldsymbol{\epsilon} $

\EndFor

\State \Return $\mathbf{x}_1'$
\end{algorithmic}
\end{algorithm}



\begin{figure}[b]
    \centering

    \subfloat[][]{%
        \includegraphics[width=0.112\columnwidth,page=1]{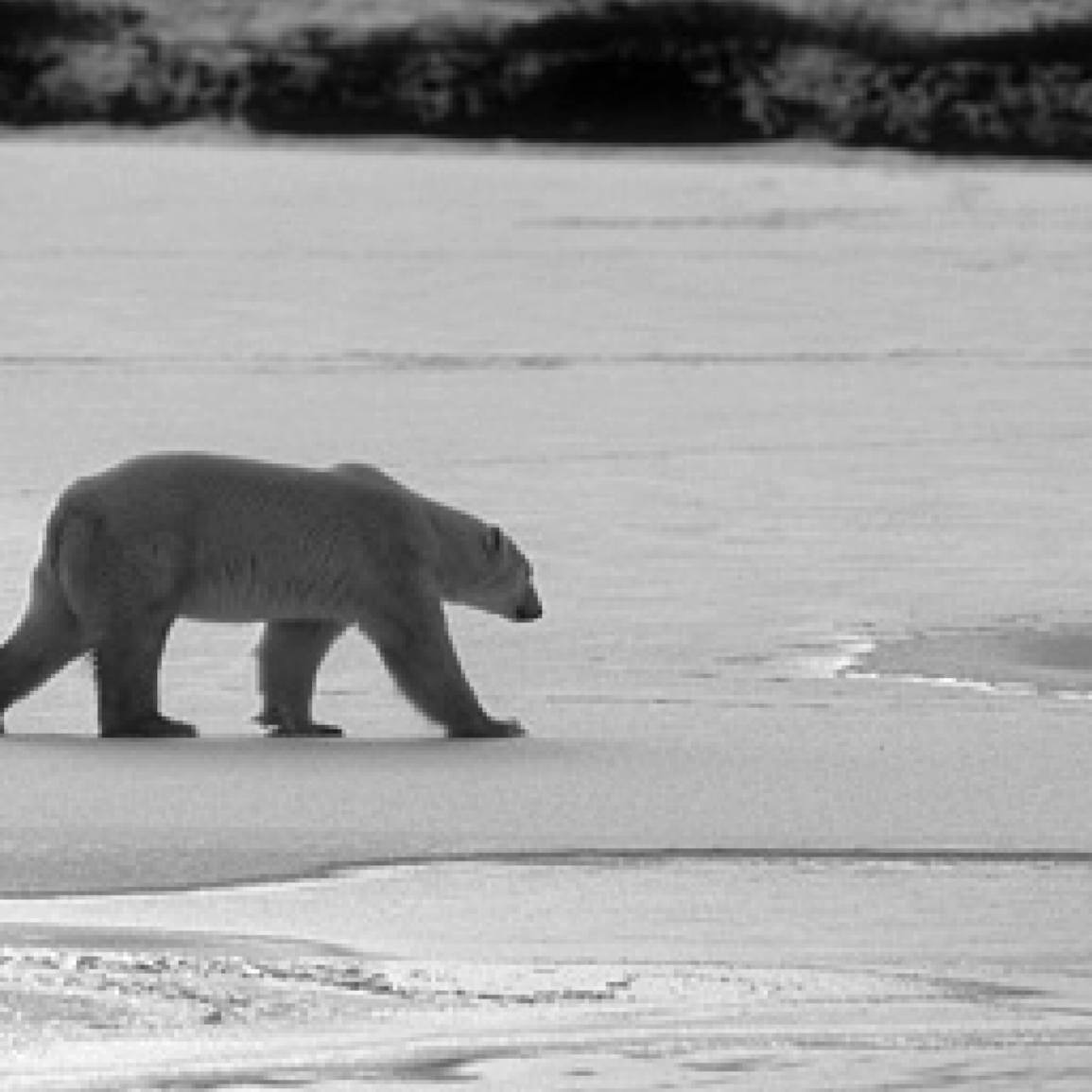}
    }
    \subfloat[][]{%
        \includegraphics[width=0.112\columnwidth,page=1]{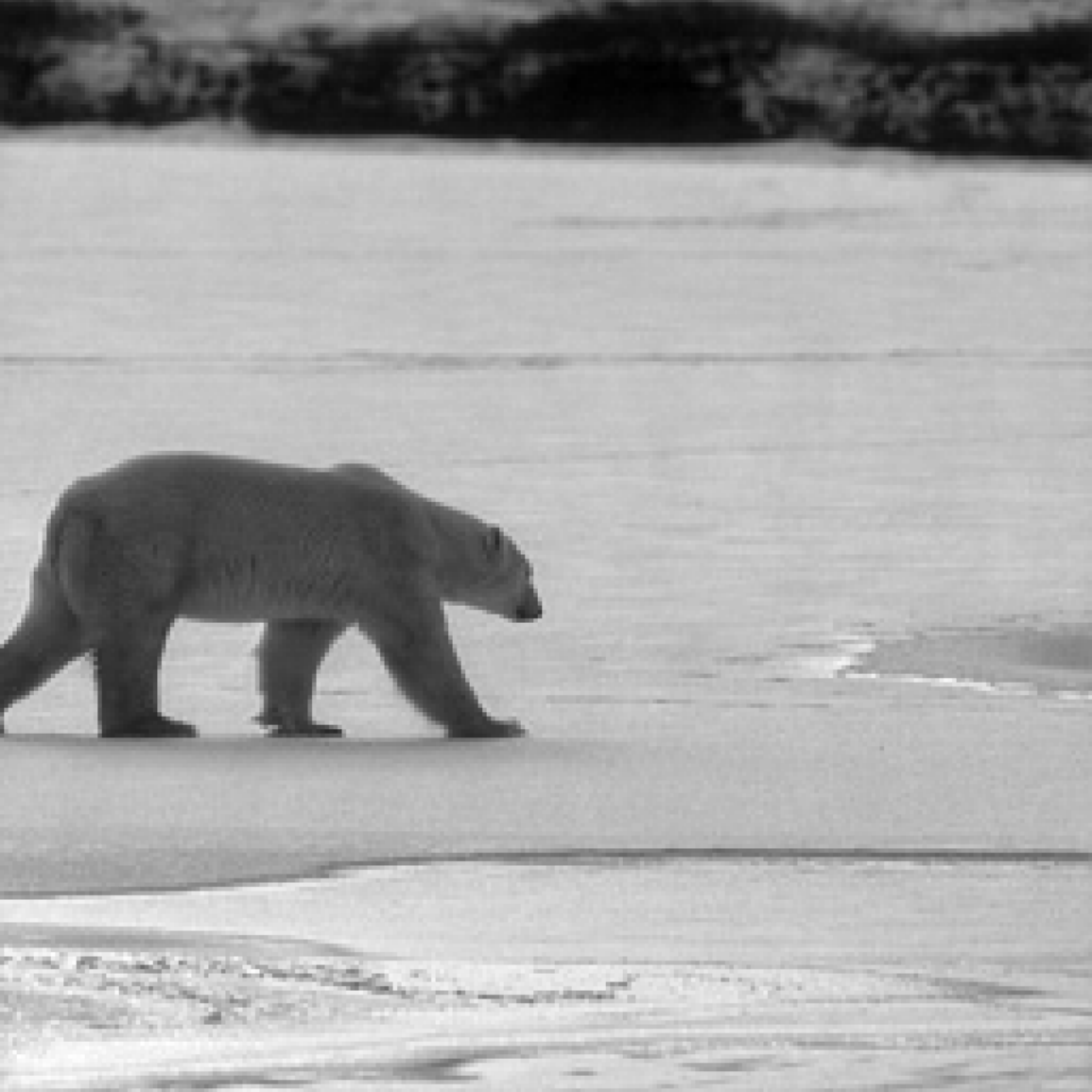}
    }
    \subfloat[][]{%
        \includegraphics[width=0.112\columnwidth,page=1]{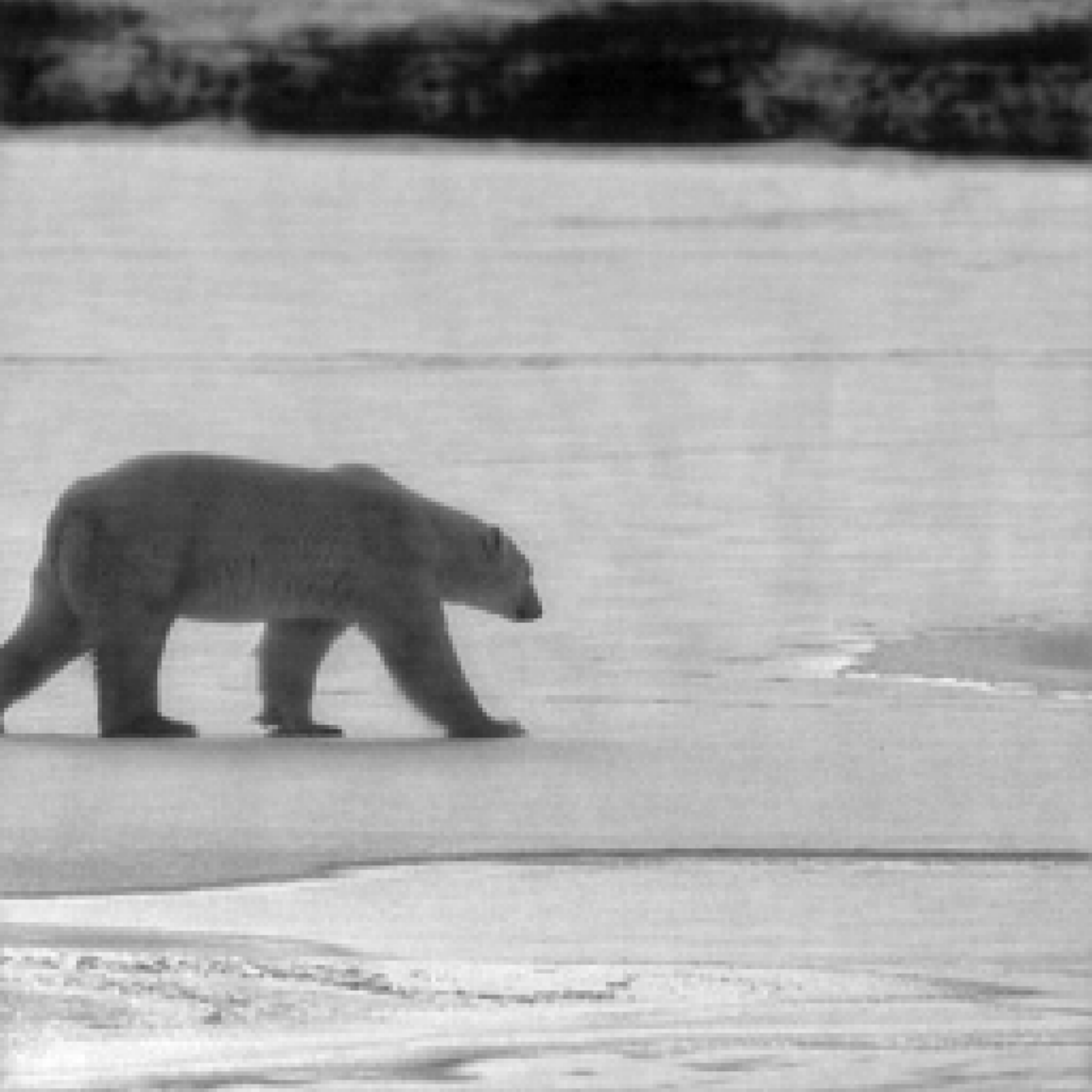}
    }
    \subfloat[][]{%
        \includegraphics[width=0.112\columnwidth,page=1]{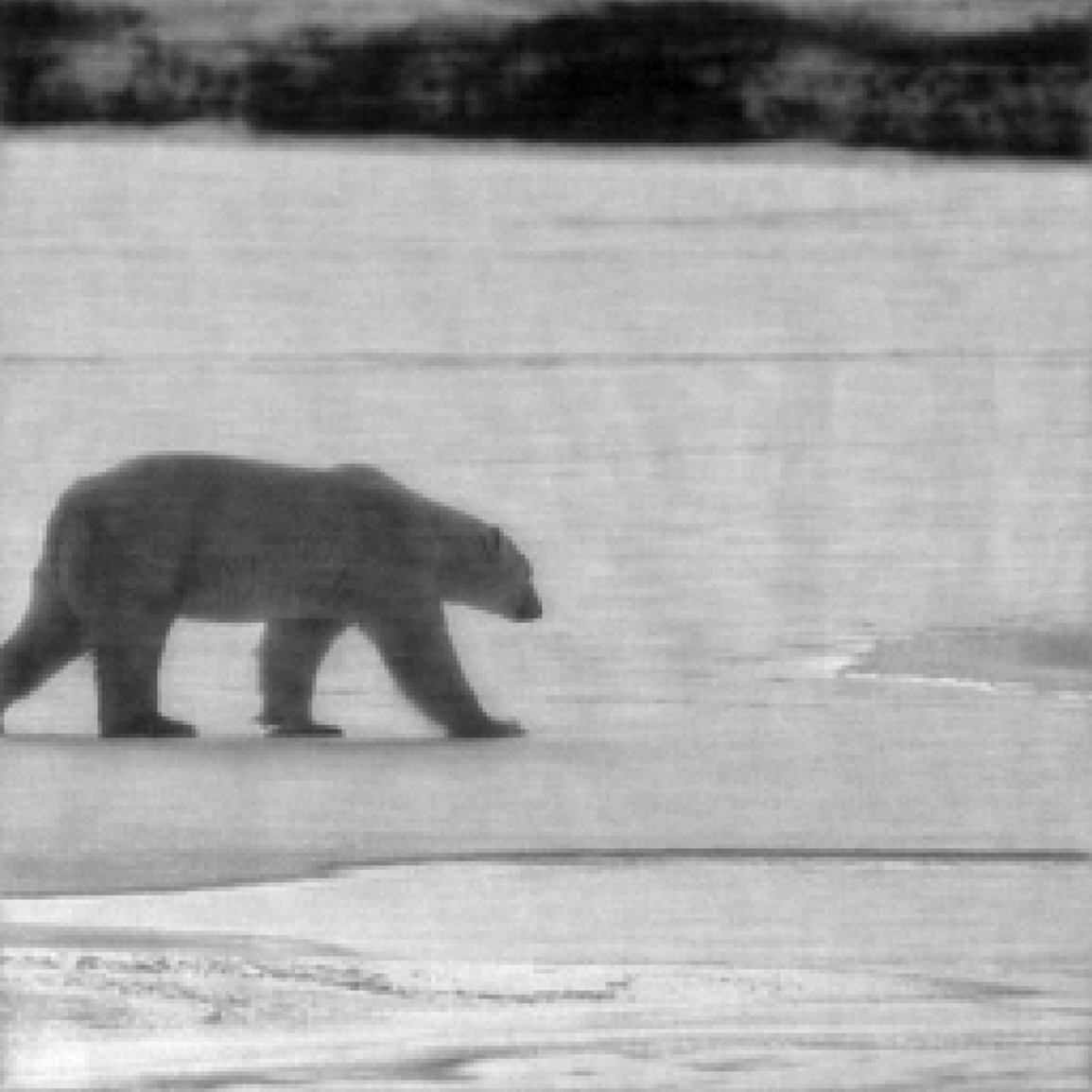}
    }
    \subfloat[][]{%
        \includegraphics[width=0.112\columnwidth,page=1]{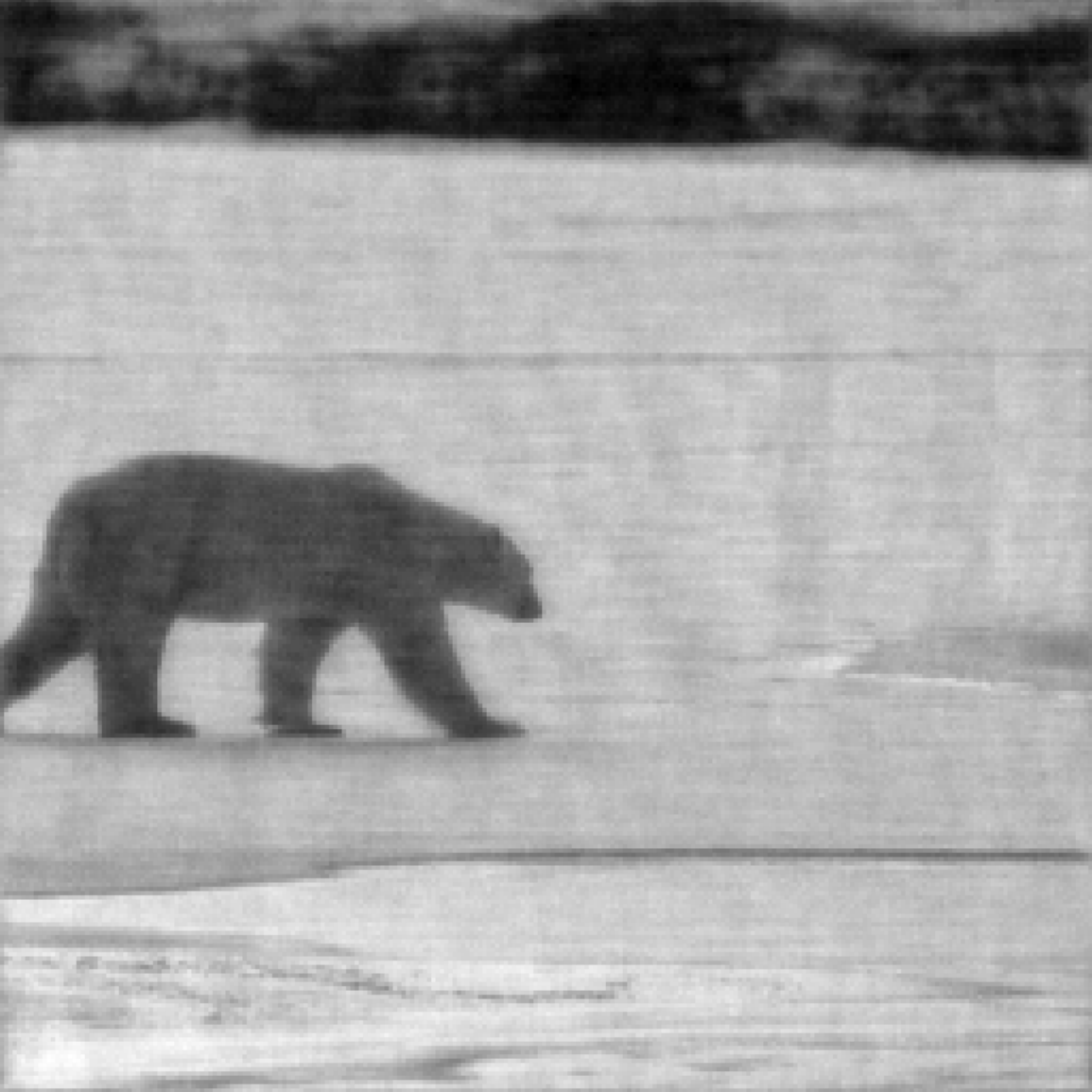}
    }
    \subfloat[][]{%
        \includegraphics[width=0.112\columnwidth,page=1]{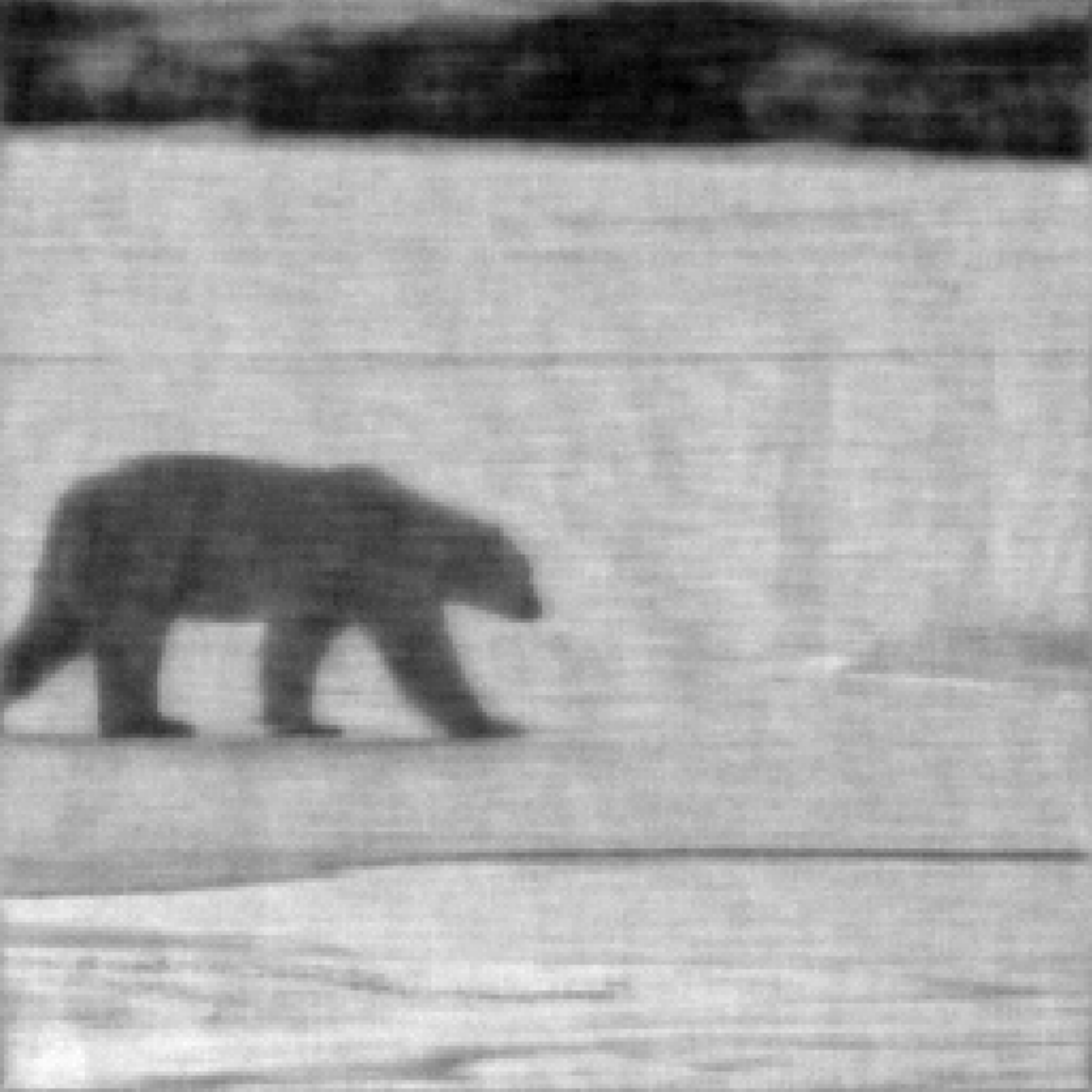}
    }
    \subfloat[][]{%
        \includegraphics[width=0.112\columnwidth,page=1]{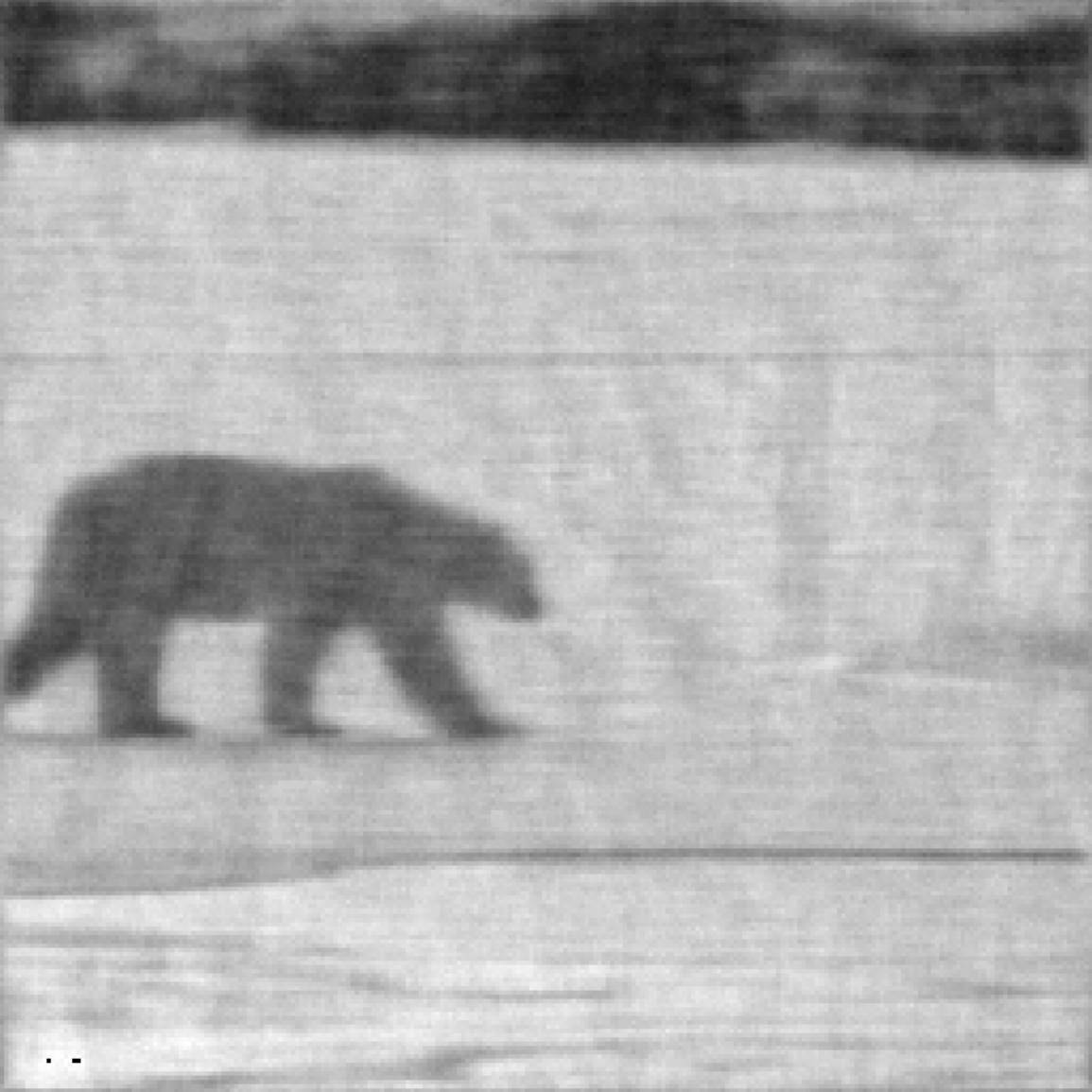}
    }
    \subfloat[][]{%
        \includegraphics[width=0.112\columnwidth,page=1]{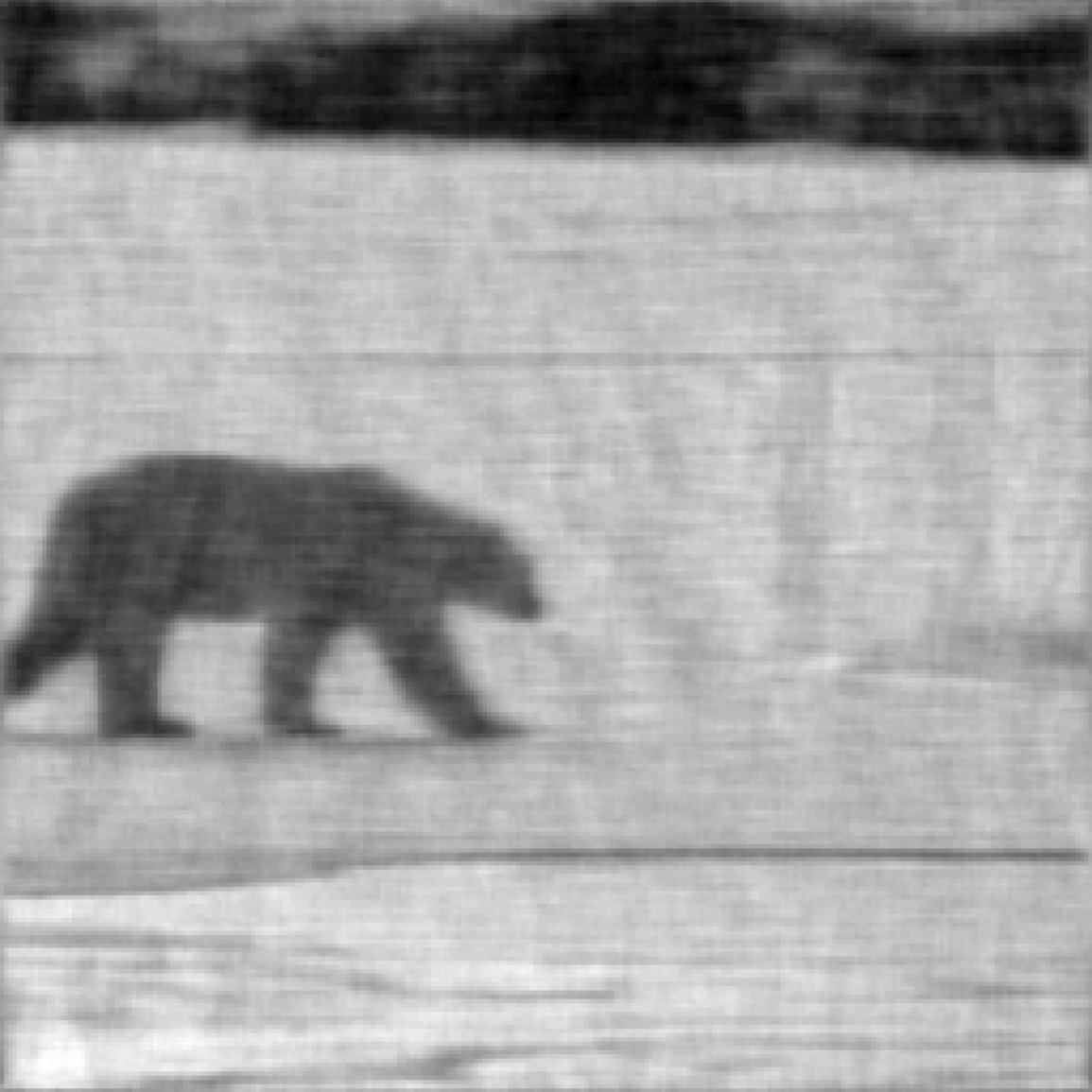}
    }

    \caption{An example of the defined gradual degradation process used during training. The timestep increases from left to right. The rightmost image ($t=7$) corresponds to the output of the initialization procedure, and the leftmost image ($t=0$) corresponds to the clean image.}
    \label{fig:indinonononoiseprocess}
\end{figure}

For training our approach, we follow the InDI training strategy, which is based on a carefully defined degradation/noising process. An example of the gradual degradation used during training can be seen in Fig. \ref{fig:indinonononoiseprocess}. Training the refinement model at any timestep requires the output from the previous step. In this training framework, simulating the previous step's refined output is straightforward due to the structured degradation process, enabling the model to learn effectively from the transition between noise and clean images. This simulated output used in training aligns with the iterative refinement output during testing due to the mathematical properties of the InDI degradation process. This alignment is essential for stable and useful training. It not only enhances the model's performance by applying the learned denoising capabilities to progressively improve the quality of the output but also makes the training computationally efficient and fast.

\begin{figure}[t]
\centering\includegraphics[width=\columnwidth, page=1]{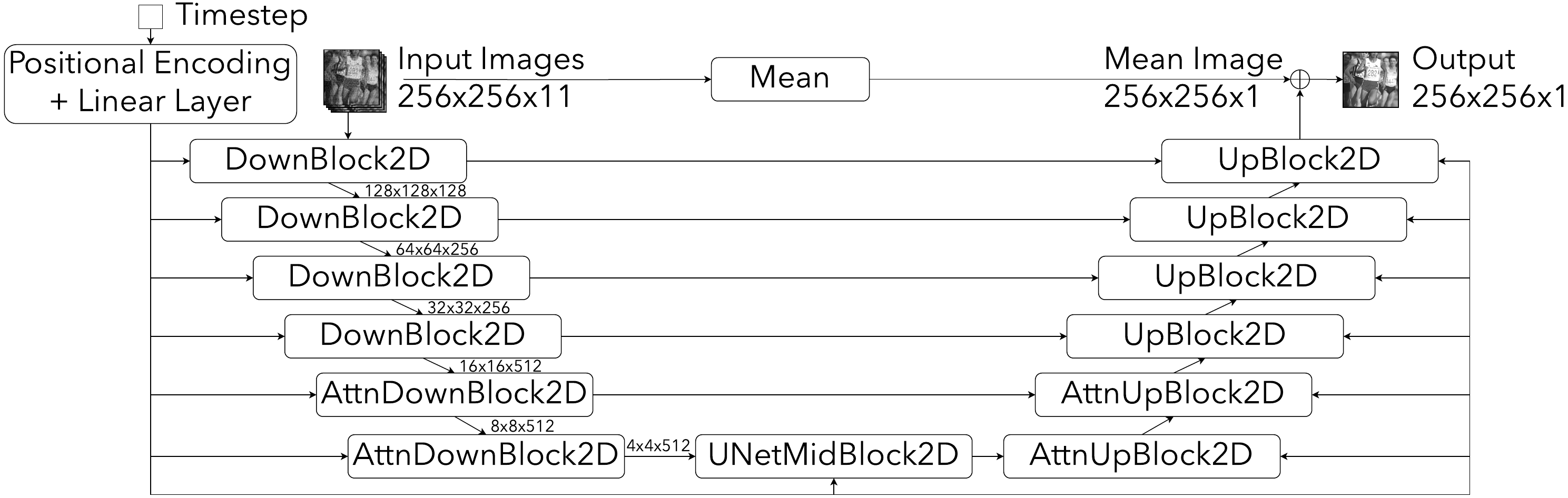}
\caption{Architecture of the used denoiser with multiple input images and a timestep input, producing a single denoised image. The network follows a UNet structure. Following the conventions defined in \cite{diffusers}, this architecture first utilizes a downsampling path composed of DownBlock2D layers that progressively reduce spatial resolution while increasing the number of channels and enhancing feature extraction. AttnDownBlock2D layers incorporate attention to focus on relevant spatial information. At the bottleneck, the UNetMidBlock2D refines the lowest resolution features before upsampling begins. The upsampling path mirrors the downsampling, with UpBlock2D and AttnUpBlock2D layers restoring spatial resolution with attention-based refinement. Skip connections between corresponding downsampling and upsampling blocks help preserve spatial information across resolutions. The final denoised image is obtained by subtracting the predicted noise (residual) from the mean of the input images.}
\label{fig:indinononounetarchitecture}
\end{figure}

As the denoising component of our approach, we employ a customized UNet architecture, as depicted in Fig. \ref{fig:indinononounetarchitecture}. The UNet architecture is well-established in image restoration tasks, known for its ability to capture both local and global features via a combination of downsampling and upsampling paths.
Our implementation of UNet operates on a set of $k+1$ input images, which include the current best estimate in the InDI refinement process and $k$ crude estimates from the initialization stage. The model receives these input images as distinct channels within a single input tensor, providing the model with richer contextual information to guide its refinement process. In addition to these images, a timestep value is also provided as an auxiliary input.
This timestep is associated with the noise level at each denoising step and is encoded using a positional encoding followed by a linear transformation. By conditioning the network on the noise level via the timestep, the model can adapt its behavior to the varying levels of noise in the images as the iterations progress.

The UNet denoiser operates by estimating the residual relative to the mean of the $k+1$ input images, rather than directly predicting the clean image. The residual represents the discrepancy between the noisy input and the desired clean output. By iteratively refining this residual, the network progressively improves the reconstruction, focusing on reducing the remaining noise while preserving essential image details. This approach is computationally efficient, as the network only needs to focus on small corrections at each iteration, ensuring that even subtle features are preserved and refined.

Our customized UNet architecture contains several enhancements, most notably the integration of attention mechanisms within the convolutional blocks. These attention mechanisms enable the network to selectively focus on the most relevant parts of the input images, enhancing its ability to capture intricate details and effectively remove noise and artifacts. The attention layers, applied in both the downsampling and upsampling paths, help the network concentrate on important features, particularly in regions where the noise or artifacts are more prevalent. This selective focus is critical in phase retrieval tasks, where noise levels can vary significantly across different regions of the image.

Additionally, the use of attention mechanisms helps preserve high-frequency details that are crucial for accurate reconstruction. By refining both coarse and detailed structures, the attention layers ensure that important features, such as edges and textures, are retained during the denoising process. These attention modules work alongside the convolutional layers to enhance the network's ability to remove noise without sacrificing image quality. Overall, the integration of attention mechanisms significantly boosts the model's capacity to handle complex noise patterns, ensuring a more accurate and visually appealing output.

\subsection{Initialization Stage}

The challenge of phase retrieval is intensified by the inherent nonlinearity and nonconvexity of the problem. These characteristics make the solution methods highly sensitive to how the reconstruction process is initialized. In our method, to mitigate these challenges and improve the robustness of the reconstruction, we adopt a sophisticated initialization strategy that builds upon the principles described in \cite{pmlr-v80-metzler18a}. This strategy involves a hybrid approach that combines the Hybrid Input-Output (HIO) method with Error Reduction (ER) and also involves an acceleration mechanism.

Initially, the HIO method is employed using $m$ different random initializations. This step explores various potential starting points in the solution space, each initialized using a different random phase. For each different initialization, the HIO algorithm is run for a small fixed number of iterations (denoted as \( s \)), which allows each initialization to evolve enough without significant computational overhead. This initial exploration aims to identify promising regions in the search space and is highly parallelizable. After this initial exploration, each result is evaluated by computing the residual \(\|\mathbf{y}-|\mathbf{A} \mathbf{x}|\|_2^{2}\). The $k$ results yielding the lowest residuals are selected for further refinement.

These chosen estimates undergo additional HIO and ER processing for an extended number of iterations \( n' \), to enhance the fidelity of the reconstructions. This additional processing is structured in cycles, where the reconstruction alternates between applying the HIO constraint and the ER constraint. Specifically, for a fixed number of iterations, the HIO algorithm is applied to improve the estimate while encouraging consistency with the measured intensity data. This is followed by a few iterations of the ER algorithm, which directly minimizes the error in the image domain. The process of alternating between these two constraints helps to prevent the reconstruction from getting stuck in local minima by balancing aggressive exploration of the solution space (via HIO) with fine-tuned refinement (via ER). This cyclic alternation continues until convergence is achieved or the maximum iteration limit is reached, resulting in a more accurate reconstruction.

\begin{figure}[tb!]
\centering\includegraphics[width=\columnwidth, page=1]{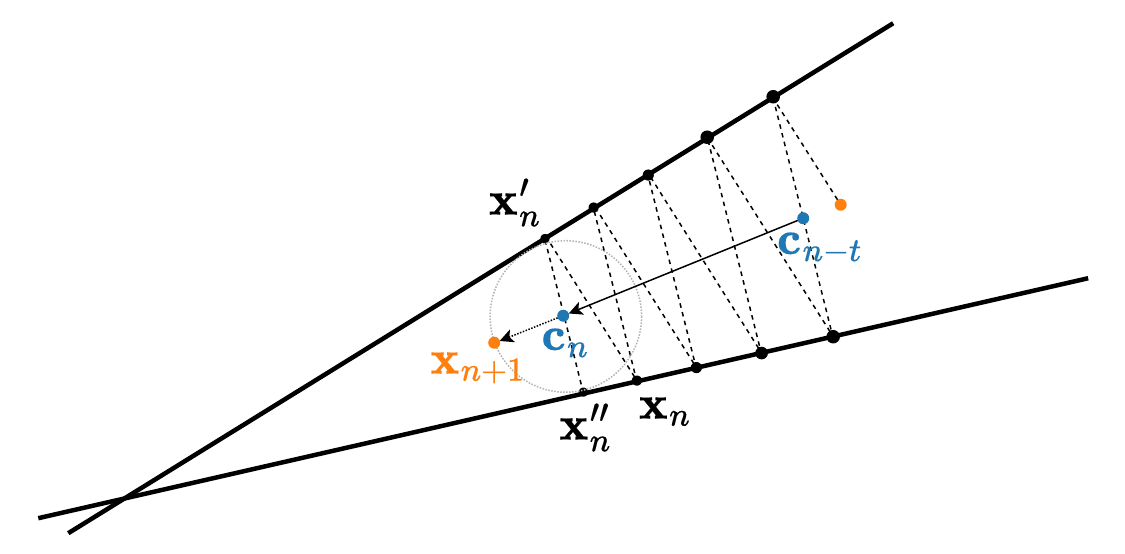}
\caption{Geometric interpretation of the acceleration mechanism used during the ER phase in the initialization procedure. The mechanism employs momentum-like updates between successive projections to enhance convergence speed and stability.}
\label{fig:indinononoaccelerateder}
\end{figure}

Notably, during the ER phases, an acceleration step is incorporated every few iterations, as shown in Fig. \ref{fig:indinononoaccelerateder} and Algorithm \ref{alg:accelerateder_for}.
This acceleration mechanism is designed to enhance convergence speed by dynamically adjusting the current estimate. It leverages a convex combination of the current and previous estimates, moderated by a scaling factor $\zeta$, which balances the contribution of these two estimates. This method aims to exploit information from previous iterations to maintain momentum, allowing the algorithm to escape stagnation points and avoid getting trapped in local minima. By integrating this momentum-like effect, the acceleration step propels the reconstruction process towards the ground truth more effectively, thus reducing the number of iterations required for convergence. Empirically, this approach has been observed to significantly speed up the convergence rate, enhancing both computational efficiency and reconstruction quality.
By incorporating an additional dynamic adjustment step, the proposed accelerated ER (AER) algorithm leverages geometric insights from successive iterations to guide the reconstruction more effectively.

At each acceleration step, the algorithm computes the difference between two successive projections onto measurement and space domain constraints, \(\mathbf{x}_n'\) and \(\mathbf{x}_n''\), which captures the current "radius" of the oscillation between projections. This difference forms an intermediate point \(\mathbf{c}_n\), representing a more balanced position between the two projections. Comparing \(\mathbf{c}_n\) with the corresponding point from earlier iterations \(\mathbf{c}_{n-t}\), the algorithm extracts a directional vector \(\mathbf{a}\), which estimates the trajectory towards the true solution.

The scalar \(r = \frac{1}{2} \|\mathbf{x}_n' - \mathbf{x}_n''\|\) serves as a local estimate of the progression distance in the current iteration. Using the scaling factor \(\zeta\), this radius is then employed to adjust the next estimate \(\mathbf{x}_{n+1}\) with the goal of accelerating convergence.

\begin{algorithm}[t]
\caption{Proposed accelerated ER (AER) algorithm}\label{alg:accelerateder_for}
\begin{algorithmic}[1]
\For{$ n = 1 $ to $K$}
    \State $\mathbf{x}_n' \gets \mathcal{P}_F \mathbf{x}_n$
    \State $\mathbf{x}_{n}'' \gets \mathcal{P}_S \mathbf{x}_n'$\\\vspace{-5pt}
    \If{$n \equiv -1 \pmod{t}$}
        \State $\mathbf{c}_n \gets \frac{1}{2} (\mathbf{x}_n' + \mathbf{x}_{n}'')$
        
        \State $\mathbf{a} \gets \frac{\mathbf{c}_n - \mathbf{c}_{n-t}}{\| \mathbf{c}_n - \mathbf{c}_{n-t} \|}$
        
        \State $r \gets \frac{1}{2} \|\mathbf{x}_n' - \mathbf{x}_{n}''\|$
        \State $\mathbf{x}_{n+1} \gets \mathbf{c}_n + \zeta r \mathbf{a}$
    \Else
        \State $\mathbf{x}_{n+1} \gets \mathbf{x}_{n}''$
    \EndIf
\EndFor
\end{algorithmic}
\end{algorithm}
Geometrically, the radius \(r\) provides an upper bound on the distance to the true solution, ensuring that updates are made in the direction of the largest progression. This adjustment helps avoid overshooting while maintaining progress towards the global optimum. In practice, this momentum-like mechanism helps escape local traps and prevents stagnation, resulting in faster convergence and improved reconstruction quality.

Empirically, this initialization procedure with acceleration and iterative refinement approach has demonstrated superior performance, particularly in terms of achieving lower residuals and higher image quality. 
Hence, our initialization framework not only exploits established algorithms but also innovates by integrating a tailored acceleration technique during the ER phases.
The inclusion of an acceleration mechanism during the ER steps further enhances the efficiency of the reconstruction process, allowing for faster convergence while maintaining the integrity of the reconstructed image as they reliably reduce the error in each iteration.
This significantly improves the robustness and effectiveness of the phase retrieval process, making it well-suited for complex imaging scenarios where traditional methods struggle.


\subsection{Geometric Self-Ensemble and Aggregation Scheme}

Since our overall pipeline, including the initialization stage, incorporates stochastic elements, such as random initial phases and Gaussian noise, it naturally produces different outputs from the same intensity measurement. This variability offers an opportunity to generate multiple reconstructions by running the pipeline repeatedly. By aggregating these outputs, we can combine the strengths of each individual reconstruction, mitigating the weaknesses of any single output. This strategy improves robustness and reconstruction quality. The aggregation process leverages the stochastic nature of our pipeline to produce a more accurate and reliable final reconstruction.

\begin{figure}[tb!]
\centering\includegraphics[width=\columnwidth, page=1]{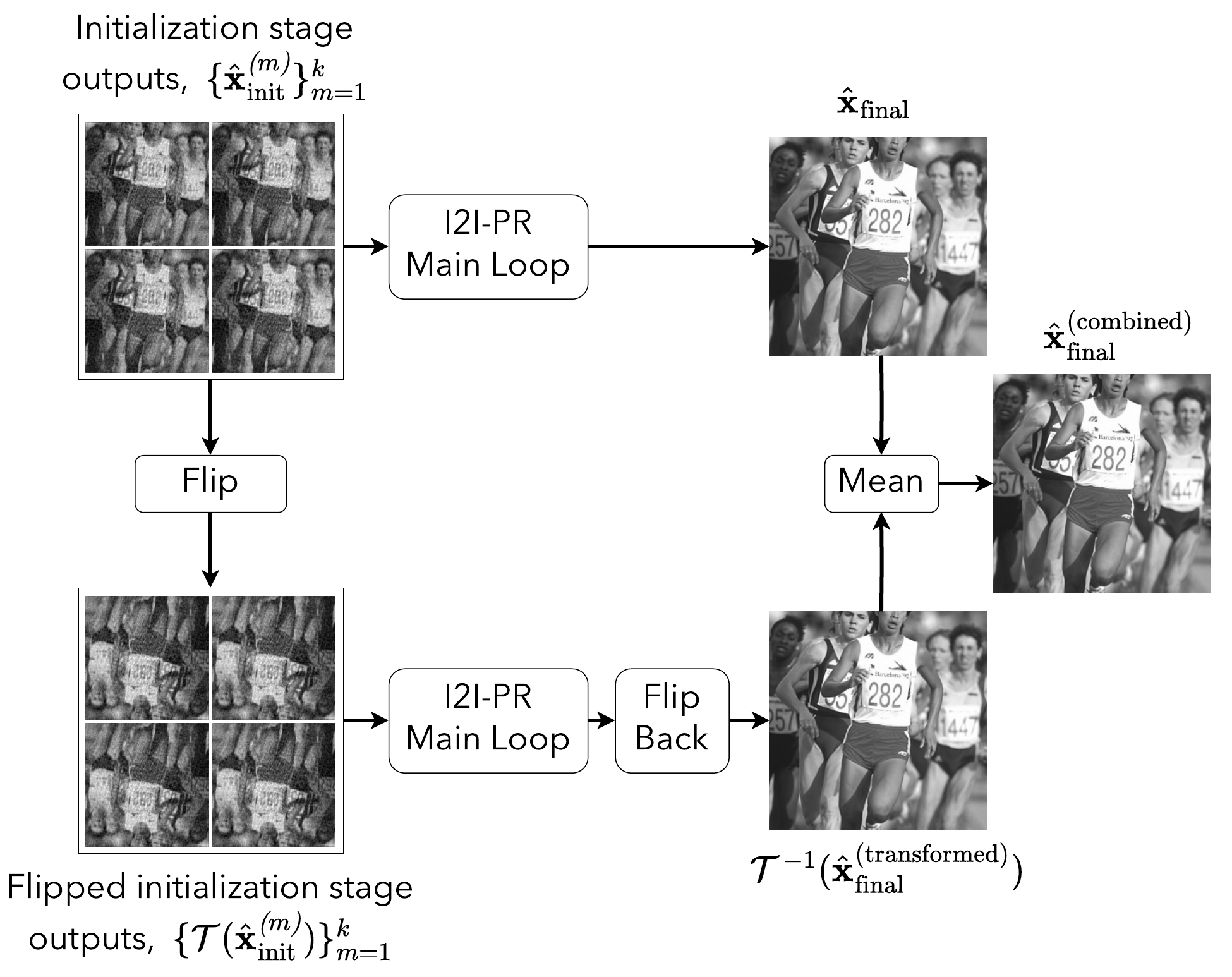}
\caption{Augmentation of the initialization outputs with an equivariant transform.}
\label{fig:indinononoaugmentaed}
\end{figure}

However, running the entire pipeline, including the initialization stage, multiple times can be time-consuming. To address this, we can augment the initialization output by applying a simple transform and reuse it instead of generating a completely new initialization. This test-time augmentation method, known as geometric self-ensemble~\cite{zhang2021plug}, allows us to skip the expensive initialization stage and only rerun the iterative refinement, significantly reducing the computational cost while still introducing diversity in the final outputs. Since the main loop relies on intensity measurements, the applied augmentation should be equivariant to these measurements, ensuring the augmented output remains consistent with the given measurement data.
Thus, we can process the initialization outputs $\{ \hat{\mathbf{x}}_{\text{init}}^{\textit{(m)}} \}_{m = 1}^k$ with an equivariant transform $\mathcal{T}$, such as flipping,
to easily produce a different set of initializations $\{ \mathcal{T}( \hat{\mathbf{x}}_{\text{init}}^{\textit{(m)}} ) \}_{m = 1}^k$. Then, we iteratively refine these two different initialization outputs and get two different final outputs. Then, we combine the two resulting final outputs through simple averaging, i.e., $ \hat{\mathbf{x}}_{\text{final}}^{\text{(combined)}} = \frac{1}{2} \left(\hat{\mathbf{x}}_{\text{final}} + \mathcal{T}^{-1} (\hat{\mathbf{x}}_{\text{final}}^{\text{(transformed)}}) \right) $, as illustrated in Fig. \ref{fig:indinononoaugmentaed}.

We produce $p$ different such combined self-ensemble results, $\hat{\mathbf{x}}_{\text{final}}^{\text{(combined)}}$, by starting our algorithm from scratch with different initialization. Because our algorithm has stochastic components, including the random initial phase in the initialization stage and additive Gaussian noise in the iterative refinement stage, each combined result is different.
We then again aggregate to obtain the final result. 
As the result of such an aggregation scheme, effectively, we combine $2 p$ samples $\{ \hat{\mathbf{x}}_{\text{final}}^{\text{(q)}} \}_{q=1}^{2p}$ from the posterior distribution $p(\mathbf{x} | \mathbf{y})$. As $p$ grows, the ensemble average, $\bar{\mathbf{x}}_{\text{final}} = \frac{1}{2p}\sum_q \hat{\mathbf{x}}_{\text{final}}^{\text{(q)}}$, converges to the MMSE estimate, and we expect to see better distortion metrics.

\section{Experiments and Results}
\label{sec:resultsindi}

To evaluate the performance of our method, we conduct numerical simulations using a large image dataset. We analyze robustness to noise, generalization capacity, and computational cost. We compare the reconstruction performance against both classical and state-of-the-art phase retrieval methods. For both testing and training stages, the noisy Fourier measurements are simulated using Eq. \ref{eq:eq1general}, with the average SNR values presented in Table \ref{table:indinononoresultsqua} (where SNR $= 10 \log(\left\Vert \vert \mathbf{Fx} \vert^2 \right\Vert_2 / \left\Vert \mathbf{y^2-\vert Fx \vert^2} \right\Vert_2)$). To ensure uniqueness (aside from trivial ambiguities), we use an oversampled discrete Fourier transform (DFT) with the oversampling rate of $m = 4n$, as suggested in~\cite{hayes1982}.  

\subsection{Training Details}

For the training phase, exclusively natural images are utilized. The training dataset comprises $44,000$ natural images, including $200$ training and $100$ validation images from the Berkeley segmentation dataset (BSD)~\cite{martin2001bsd}, $41,400$ images selected from the ImageNet dataset~\cite{deng2009imagenet,zhang2017learning}, and $2,300$ images randomly chosen from the Waterloo Exploration Database~\cite{ma2017waterloo}.

Decoupled weight decline regularization \cite{Loshchilov2017DecoupledWD}  together with cosine annealing with linear warming \cite{Loshchilov2016SGDRSG} is used for the training to minimize MSE loss.
The developed method is implemented by PyTorch and tested on an 
NVIDIA A100 80GB PCIe GPU. The total training time is about 60 hours (27 epochs).

\subsection{Testing and Implementation Details}

During the testing phase, both natural and unnatural images are employed. This allows us to evaluate the performance of the methods not only for real-world scenarios (natural images) but also for unseen synthetic samples (unnatural images). That is, despite the training being conducted solely with natural images, the developed method is evaluated on both natural and unnatural images to assess its generalization capabilities.
This test dataset, previously used in \cite{Isil:19, Isil:20, isil2024deep}, contains $236$ images, which include $230$ natural and $6$ synthetic images. Specifically, the dataset consists of $200$ test images from BSD, $24$ images from the Kodak dataset~\cite{franzen}, and $6$ natural and $6$ synthetic images taken from \cite{pmlr-v80-metzler18a}. The synthetic images are from scanning electron microscopes and telescopes. All images have pixel values ranging from $0$ to $255$ and are of size $256 \times 256$.

In the initialization stage,
the HIO method is first run with $m=100$ different random initializations for $s=50$ iterations. Then, $k=10$ reconstructions 
with the lowest residuals
is used for AER+HIO run for $n=1700$ iterations with $\zeta = 0.6$. 
Thus, as the output of this initialization procedure, $k=10$ multiple outputs are generated from the best $k=10$ initializations with the lowest residuals among the $m=100$ different random initializations. 
In the iterative stage, the hyperparameters are set as follows: \(K = 5\), \(\beta = 0.9\), and \(\sigma_i = 1 \, \forall i\). The learnable vector \(\boldsymbol{\lambda} \in \mathbb{R}^T\) is initialized with logarithmically increasing values from \(10^{-0.64}\) to \(10^{-0.10}\).



\subsection{Performance Analysis}

\renewcommand{\arraystretch}{1.12}
\begin{table*}[tb!]
	\centering
	\caption{Average reconstruction performances for $236$ test images across $5$ Monte Carlo runs.}
	    \begin{adjustbox}{width=\textwidth}
     
		\begin{tabu}{cccccccc}
                
			\tabucline[1.3pt]{\hline-to}
			
			$\alpha = 2$ & \multicolumn{3}{c}{Avg. PSNR (dB) $\uparrow$}& \multicolumn{3}{c}{Avg. SSIM $\uparrow$}& Avg. runtime (sec.) $\downarrow$\\
			(Avg. SNR: 33.24
dB)&Overall&Natural&Unnatural&Overall&Natural&Unnatural&\\			
			\hline
HIO \cite{fienup1982comparison} & 19.79&19.73&21.92&0.50&0.50&0.49&0.25\\

prDeep \cite{pmlr-v80-metzler18a} & 23.45&23.49&21.72&0.65&0.66&0.58&59.32\\	

DIR \cite{Isil:19} & 23.61&23.60&24.02&0.72&0.72&0.73&21.59\\ 

PnP-HIO \cite{isil2024deep} & 24.87&24.86&25.56&0.74&0.74&0.74&24.11\\ 


Initialization procedure& 21.12&21.02&24.82&0.55&0.55&0.58&0.91\\
I2I-PR ($T=4,$ no aggregation)& 28.59&28.65&26.39&0.79&0.80&0.67&1.10\\


\tabucline[1.3pt]{\hline-to}

$\alpha = 3$ & \multicolumn{3}{c}{Avg. PSNR (dB) $\uparrow$}& \multicolumn{3}{c}{Avg. SSIM $\uparrow$}& Avg. runtime (sec.) $\downarrow$\\
(Avg. SNR: 31.53
dB)&Overall&Natural&Unnatural&Overall&Natural&Unnatural&\\			
\hline
HIO \cite{fienup1982comparison} & 18.92&18.89&20.34&0.43&0.43&0.43&0.27\\

prDeep \cite{pmlr-v80-metzler18a} & 22.06&22.09&20.91& 0.59&0.59&0.54&59.41\\	
   
DIR \cite{Isil:19} & 22.87&22.85&23.50&0.68&0.68&0.71&21.72\\ 

PnP-HIO \cite{isil2024deep} & 23.92&23.92&23.98&0.70&0.70&0.69&24.35\\ 


Initialization procedure& 20.17&20.12&22.09&0.51&0.51&0.54&0.90\\

I2I-PR ($T=4,$ no aggregation)& 26.78&26.85&24.18&0.73&0.73&0.61&1.11\\


   \tabucline[1.3pt]{\hline-to}

$\alpha = 4$ & \multicolumn{3}{c}{Avg. PSNR (dB) $\uparrow$}& \multicolumn{3}{c}{Avg. SSIM $\uparrow$}& Avg. runtime (sec.) $\downarrow$\\
(Avg. SNR: 30.24
dB)&Overall&Natural&Unnatural&Overall&Natural&Unnatural&\\			
\hline
HIO \cite{fienup1982comparison} & 18.52&18.48&19.80&0.39&0.39&0.40&0.28\\	

prDeep \cite{pmlr-v80-metzler18a} &  20.69&20.70&20.38&0.53&0.53&0.51&59.68\\	

DIR \cite{Isil:19} & 21.80&21.77&22.79&0.62&0.62&0.69&21.95\\ 

PnP-HIO \cite{isil2024deep} & 22.41&22.39&23.09&0.63&0.63&0.65&24.43\\ 


Initialization procedure&19.54&19.46&20.88&0.48&0.48&0.47&0.91\\

I2I-PR ($T=4,$ no aggregation)& 25.43&25.46&24.23&0.66&0.66&0.59&1.10\\

			\hline					
	\end{tabu}
     \end{adjustbox}

\label{table:indinononoresultsqua}

\vspace{10pt}
 
 
\end{table*}
\renewcommand{\arraystretch}{1.25}

In order to evaluate our developed approach, we first compare its reconstruction accuracy against original images using the peak signal-to-noise ratio (PSNR) and the structural similarity index (SSIM)~\cite{wang2004image}. For a comprehensive comparison, results are also obtained for the same test dataset using existing methods, namely prDeep \cite{pmlr-v80-metzler18a}, the classical HIO algorithm \cite{fienup1978reconstruction}, DIR \cite{Isil:19}, and PnP-HIO \cite{Isil:20, isil2024deep}.

Table \ref{table:indinononoresultsqua} shows the average reconstruction performance for $236$ test images obtained with $5$ Monte Carlo runs and varying levels of Poisson noise ($\alpha = 2, 3, 4$). As seen, our method not only outperforms the existing methods in terms of both PSNR and SSIM but also possesses computational efficiency comparable to the HIO-based initialization procedure. The performance obtained after the initialization procedure is also listed to demonstrate the effectiveness of the iterative refinement in the developed InDI approach.

As seen in Table \ref{table:indinononoresultsqua}, our approach maintains superior reconstruction quality across various noise levels ($\alpha = 2, 3, 4$), indicating robustness to noise since the developed method is trained for the noise level of $\alpha = 3$. This demonstrates the resilience and versatility of our method.

\begin{figure}[]
    \centering

    \subfloat[][\parbox{0.31\columnwidth}{\centering (a) Ground truth}]{%
        \begin{minipage}[t]{0.31\columnwidth}
            \begin{tikzpicture}
                \node[anchor=south west,inner sep=0] (image) at (0,0) {
                    \includegraphics[width=\linewidth]{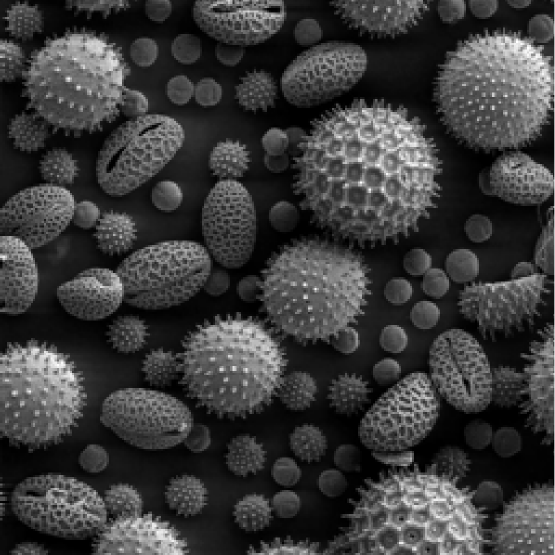}
                };
                \begin{scope}[x={(image.south east)}, y={(image.north west)}]
                    \draw[white, dashed, very thick] (0.35,0.35) rectangle (0.65,0.64);
                \end{scope}
            \end{tikzpicture}
            \vspace{2pt}
            \begin{adjustbox}{center,minipage=3.33\linewidth,trim={.35\width,.35\height,.25\width,.25\height},clip}
                \includegraphics[width=\linewidth]{figures/other_methods/117/117_0_gt_0.000_0.000_28.102.pdf}
            \end{adjustbox}
        \end{minipage}
    }
    \hfill
    \subfloat[][\parbox{0.31\columnwidth}{\centering (b) prDeep~\cite{pmlr-v80-metzler18a}\\PSNR: $19.37$,\\SSIM: $0.47$}]{%
        \begin{minipage}[t]{0.31\columnwidth}
            \includegraphics[width=\linewidth]{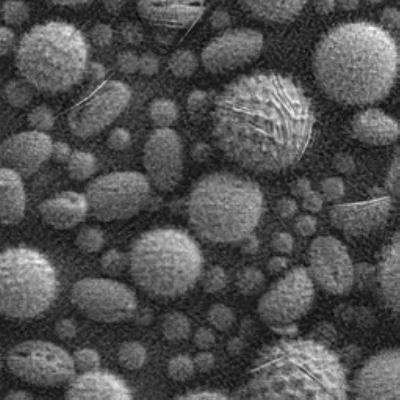}
            \vspace{2pt}
            \begin{adjustbox}{center,minipage=3.33\linewidth,trim={.35\width,.35\height,.25\width,.25\height},clip}
                \includegraphics[width=\linewidth]{figures/other_methods/cagatayao2019/204_3_1_prdeep}
            \end{adjustbox}
        \end{minipage}
    }
    \hfill
    \subfloat[][\parbox{0.31\columnwidth}{\centering (c) DIR~\cite{Isil:19}\\PSNR: $25.33$,\\SSIM: $0.67$}]{%
        \begin{minipage}[t]{0.31\columnwidth}
            \includegraphics[width=\linewidth]{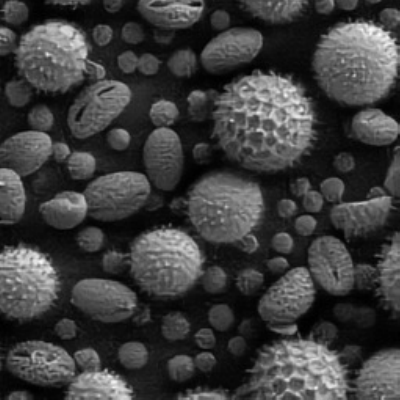}
            \vspace{2pt}
            \begin{adjustbox}{center,minipage=3.33\linewidth,trim={.35\width,.35\height,.25\width,.25\height},clip}
                \includegraphics[width=\linewidth]{figures/other_methods/cagatayao2019/204_3_1_unet2}
            \end{adjustbox}
        \end{minipage}
    }

    \subfloat[][\parbox{0.31\columnwidth}{\centering (d) PnP-HIO~\cite{isil2024deep}\\PSNR: $26.28$,\\SSIM: $0.71$}]{%
        \begin{minipage}[t]{0.31\columnwidth}
            \includegraphics[width=\linewidth]{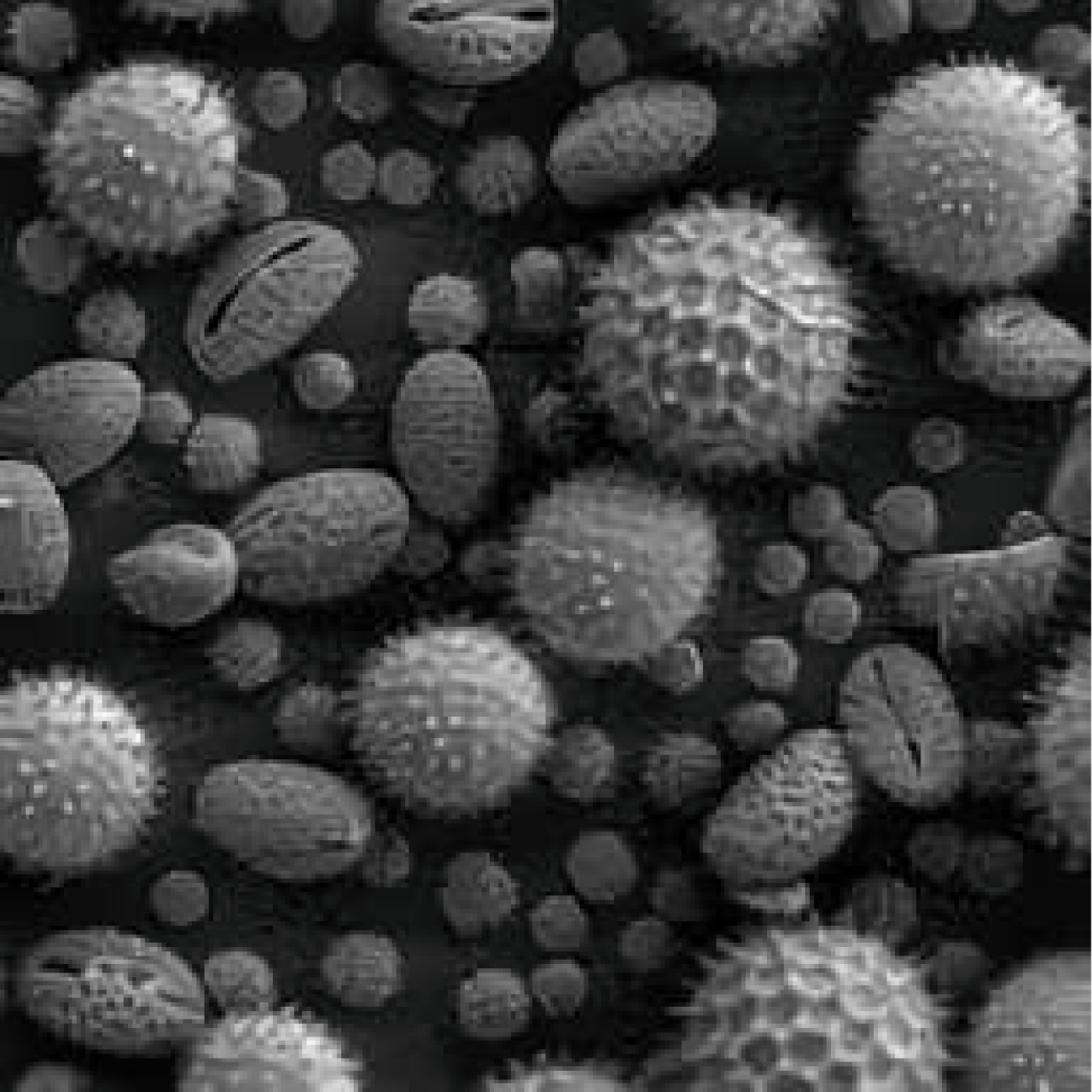}
            \vspace{2pt}
            \begin{adjustbox}{center,minipage=3.33\linewidth,trim={.35\width,.35\height,.25\width,.25\height},clip}
                \includegraphics[width=\linewidth]{figures/other_methods/117/mbwddp.pdf}
            \end{adjustbox}
        \end{minipage}
    }
    \hfill
    \subfloat[][\parbox{0.31\columnwidth}{\centering (e) I2I-PR ($T=4$,\\no aggregation)\\PSNR: $30.31$,\\SSIM: $0.90$}]{
        \begin{minipage}[t]{0.31\columnwidth}
            \includegraphics[width=\linewidth]{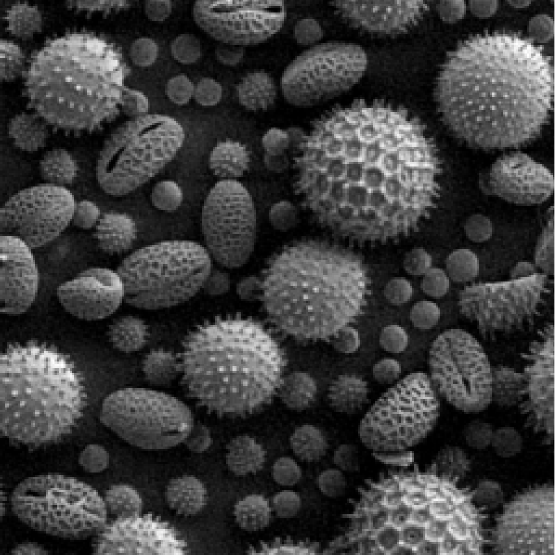}
            \vspace{2pt}
            \begin{adjustbox}{center,minipage=3.33\linewidth,trim={.35\width,.35\height,.25\width,.25\height},clip}
                \includegraphics[width=\linewidth]{figures/indipr/test_images_plot_indipr_4/117_0_final_output_developed_mean_12_30.307_0.900_27.692.pdf}
            \end{adjustbox}
        \end{minipage}
    }
    \hfill
    \subfloat[][\parbox{0.31\columnwidth}{\centering (f) I2I-PR ($T=32, 2p=24$)\\PSNR: $30.46$,\\SSIM: $0.90$}]{%
        \begin{minipage}[t]{0.31\columnwidth}
            \includegraphics[width=\linewidth]{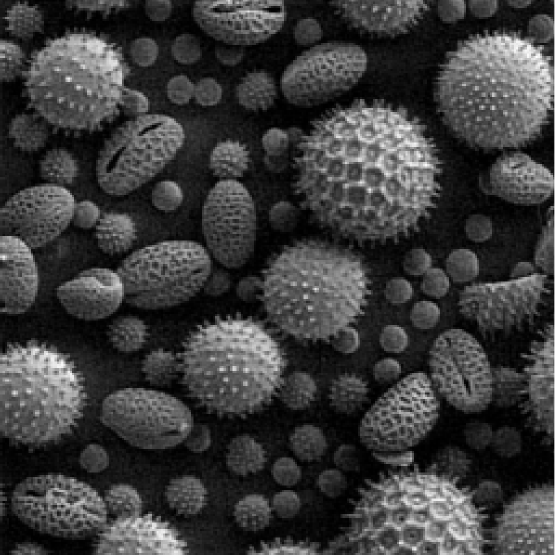}
            \vspace{2pt}
            \begin{adjustbox}{center,minipage=3.33\linewidth,trim={.35\width,.35\height,.25\width,.25\height},clip}
                \includegraphics[width=\linewidth]{figures/indipr/test_images_plot_indipr_32/117_3_final_output_developed_mean_12_30.457_0.899_28.277.pdf}
            \end{adjustbox}
        \end{minipage}
    }

    \caption{The outputs of various algorithms for the out-of-domain "Pollen" test image subjected to $\alpha=3$ noise (SNR=28.10dB).}
    \label{fig:indinononopollen}
\end{figure}

Table \ref{table:indinononoresultsqua} also presents the reconstruction performance of different methods for both natural and unnatural test images, allowing the assessment of the generalization capability of each method.
For a visual comparison, sample reconstructions of an unnatural image from the test dataset are depicted in Fig. \ref{fig:indinononopollen}.
Notably, even though our deep neural networks (DNNs) are trained exclusively on natural images, our method achieves the highest PSNR for both image types. This demonstrates the capability of our method to generalize beyond the training data to images with different characteristics and statistical properties. Despite the overall success, our method exhibits occasional shortcomings in the SSIM for unnatural images, although it consistently excels in terms of peak signal-to-noise ratio (PSNR). These discrepancies highlight potential areas for improvement, particularly in how our method handles the specific textural elements of unnatural images.

\begin{figure}[]
    \centering

    \subfloat[][\parbox{0.31\columnwidth}{\centering (a) Ground truth}]{%
        \begin{minipage}[t]{0.31\columnwidth}
            \begin{tikzpicture}
                \node[anchor=south west,inner sep=0] (image) at (0,0) {
                    \includegraphics[width=\linewidth]{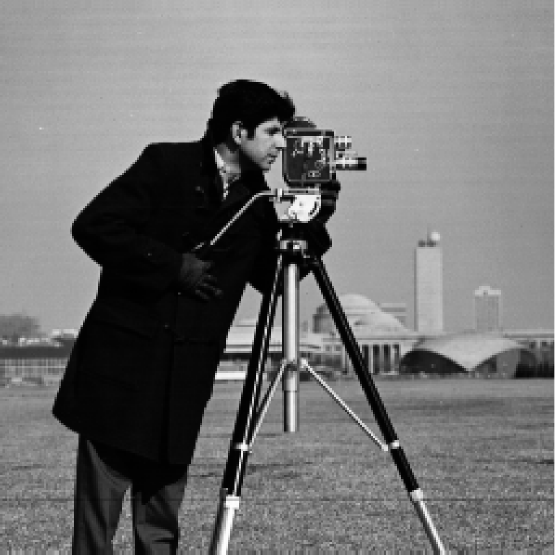}
                };
                \begin{scope}[x={(image.south east)}, y={(image.north west)}]
                    \draw[white, dashed, very thick] (0.35,0.56) rectangle (0.70,0.85);
                \end{scope}
            \end{tikzpicture}
            \vspace{2pt}
            \begin{adjustbox}{center,minipage=2.85\linewidth,trim={0.35\width, 1.1\height, 0.3\width, 0.3\height},clip}
                \includegraphics[width=\linewidth]{figures/other_methods/122/122_0_gt_0.000_0.000_31.605.pdf}
            \end{adjustbox}
        \end{minipage}
    }
    \hfill
    \subfloat[][\parbox{0.31\columnwidth}{\centering (b) prDeep~\cite{pmlr-v80-metzler18a}\\PSNR: $23.44$,\\SSIM: $0.66$}]{%
        \begin{minipage}[t]{0.31\columnwidth}
            \includegraphics[width=\linewidth]{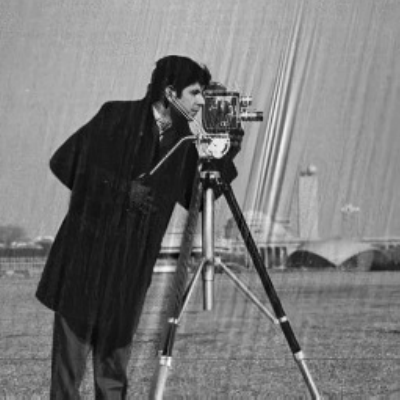}
            \vspace{2pt}
            \begin{adjustbox}{center,minipage=2.85\linewidth,trim={0.35\width, 1.1\height, 0.3\width, 0.3\height},clip}
                \includegraphics[width=\linewidth]{figures/other_methods/cagatayao2019/209_3_1_prdeep}
            \end{adjustbox}
        \end{minipage}
    }
    \hfill
    \subfloat[][\parbox{0.31\columnwidth}{\centering (c) DIR~\cite{Isil:19}\\PSNR: $20.37$,\\SSIM: $0.54$}]{%
        \begin{minipage}[t]{0.31\columnwidth}
            \includegraphics[width=\linewidth]{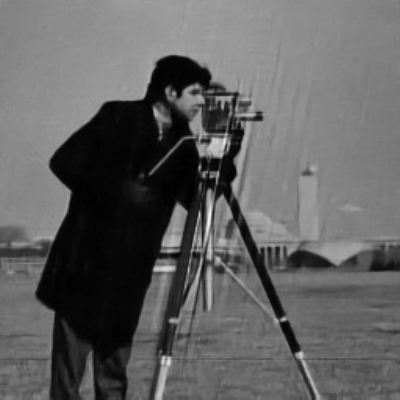}
            \vspace{2pt}
            \begin{adjustbox}{center,minipage=2.85\linewidth,trim={0.35\width, 1.1\height, 0.3\width, 0.3\height},clip}
                \includegraphics[width=\linewidth]{figures/other_methods/cagatayao2019/209_3_1_unet2}
            \end{adjustbox}
        \end{minipage}
    }

    \subfloat[][\parbox{0.31\columnwidth}{\centering (d) Initialization procedure\\PSNR: $18.96$,\\SSIM: $0.33$}]{%
        \begin{minipage}[t]{0.31\columnwidth}
            \includegraphics[width=\linewidth]{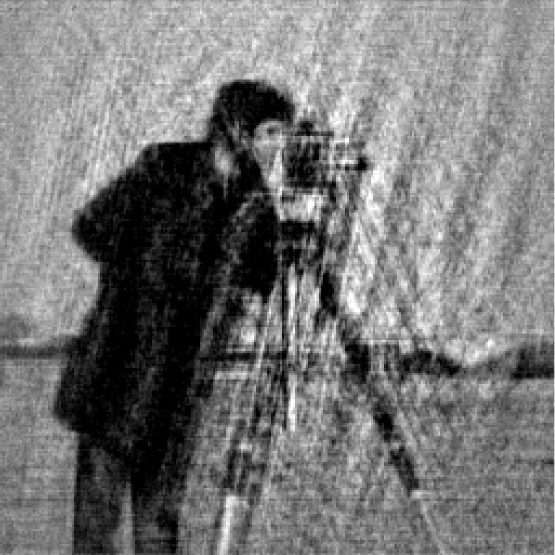}
            \vspace{2pt}
            \begin{adjustbox}{center,minipage=2.85\linewidth,trim={0.35\width, 1.1\height, 0.3\width, 0.3\height},clip}
                \includegraphics[width=\linewidth]{figures/indipr/test_images_plot_indipr_4/122_3_init_output_18.962_0.332_31.242.pdf}
            \end{adjustbox}
        \end{minipage}
    }
    \hfill
    \subfloat[][\parbox{0.31\columnwidth}{\centering (e) I2I-PR ($T=4$,\\no aggregation)\\PSNR: $27.39$,\\SSIM: $0.70$}]{%
        \begin{minipage}[t]{0.31\columnwidth}
            \includegraphics[width=\linewidth]{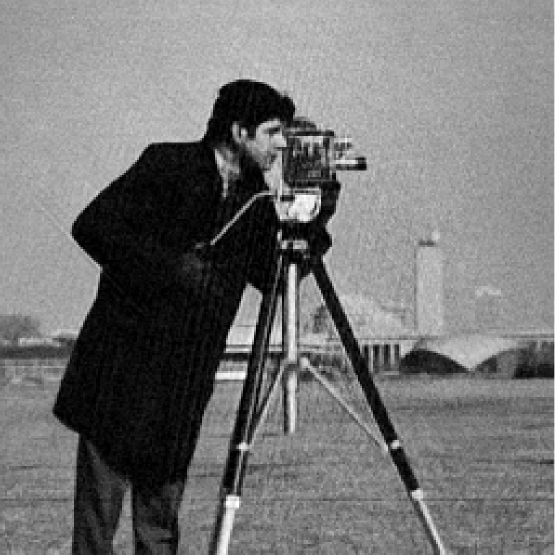}
            \vspace{2pt}
            \begin{adjustbox}{center,minipage=2.85\linewidth,trim={0.35\width, 1.1\height, 0.3\width, 0.3\height},clip}
                \includegraphics[width=\linewidth]{figures/indipr/test_images_plot_indipr_4/122_1_output_27.388_0.700_31.900.pdf}
            \end{adjustbox}
        \end{minipage}
    }
    \hfill
    \subfloat[][\parbox{0.31\columnwidth}{\centering (f) I2I-PR ($T=32, 2p=24$)\\PSNR: $28.98$,\\SSIM: $0.82$}]{%
        \begin{minipage}[t]{0.31\columnwidth}
            \includegraphics[width=\linewidth]{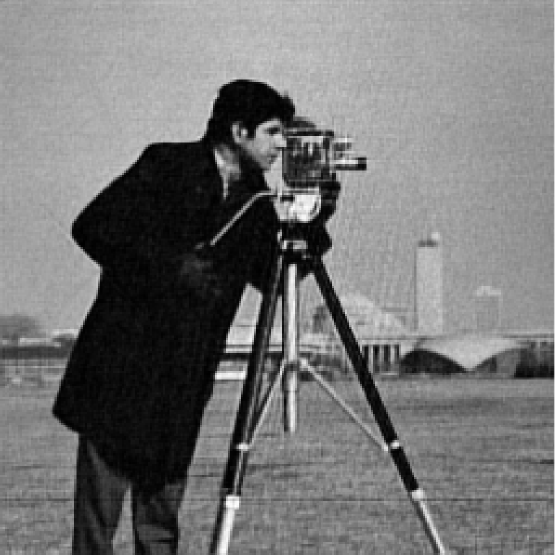}
            \vspace{2pt}
            \begin{adjustbox}{center,minipage=2.85\linewidth,trim={0.35\width, 1.1\height, 0.3\width, 0.3\height},clip}
                \includegraphics[width=\linewidth]{figures/indipr/test_images_plot_indipr_32/122_4_final_output_developed_mean_12_28.984_0.819_31.810.pdf}
            \end{adjustbox}
        \end{minipage}
    }

    \caption{The outputs of various algorithms for the "Cameraman" test image subjected to $\alpha=3$ noise (SNR=31.61dB).}
    \label{fig:indinononocameraman}
\end{figure}

Ambiguities in phase retrieval may explain the lower performance observed for unnatural images.
Note that our assumptions of realness and positivity in the image inherently mitigate the trivial global phase shift ambiguity. Moreover, in our analysis, we confront the conjugate inversion ambiguity by comparing both the original and flipped versions of the generated image with the ground truth, ensuring accurate orientation alignment of the reconstructed objects.
However, challenges exist for spatial circular shift ambiguity, particularly notable in unnatural images such as "E.Coli" and "Yeast," which do not conform well to the typical support pattern for natural images. This misalignment can lead to multiple valid reconstructions using the HIO algorithm, introducing notable ambiguities. Previous literature on phase retrieval has only sparingly discussed this issue, with few exceptions such as the studies by \cite{Goy2018LowPC, Uelwer2019PhaseRU}. While methods such as the shrinkwrap procedure \cite{marcheshrinkwrap} are known to refine support and reduce ambiguities, we choose not to implement this step in order to focus instead on the reconstruction of natural images and avoid overcomplicating the methodology.

Our decision not to use ground truth images to disambiguate circular shift ambiguity, unlike the other compared methods, explains the lower performance observed in certain cases. This inherently makes the comparison less favorable to our method, as we do not leverage the same alignment information that other methods rely on, potentially creating an unfair assessment of our model's performance in these ambiguous scenarios common with unnatural images.
The decision not to use ground truth images to disambiguate circular shift ambiguity stems from practical considerations. In real-world applications, the exact support region of the object being imaged is often unknown, making it impractical to rely on ground truth for alignment. Additionally, using the ground truth for disambiguation could introduce biases and limit the generalizability of the method to situations where such information is unavailable. By focusing on natural image reconstruction without this additional constraint, we aim to develop a more flexible approach that can better handle real-world scenarios where the support is uncertain or ill-defined.

Visually, the superiority of our approach is illustrated in Fig. \ref{fig:indinononocameraman} for a test image.
The I2I-PR approach notably excels in removing HIO artifacts and preserving image details, which is critical for high reconstruction quality.
Moreover, since we develop our method by considering the perception-distortion tradeoff, our approach effectively mitigates the common smoothing artifacts prevalent in other methods as discussed in \cite{Isil:19}. This ensures a delicate balance between minimizing distortions and preserving fine details, thus enhancing the perceptual quality of the images.

\subsection{Effect of Iteration Count}

Iteration count plays a significant role in balancing the tradeoff between image quality, perceptual similarity, and computational efficiency in our method.
To further assess how different operational settings impact reconstruction performance, we utilize perceptual quality metrics including FID \cite{fid_Heusel2017GANsTB}, LPIPS \cite{lpips_Zhang2018TheUE}, and CLIP-IQA \cite{clipiqa_Wang2022ExploringCF} in addition to the earlier used distortion metrics, i.e., PSNR and SSIM. This enables a more comprehensive analysis of image quality by evaluating not only the accuracy of pixel values but also the perceptual similarity to human vision.

Table \ref{table:indinononoresultsiterationcount} analyzes the effect of iteration count on reconstruction performance and computational efficiency for the case of $\alpha=3$ without aggregation.
It shows that a smaller number of iterations tends to yield better outcomes in terms of image quality. This is true for both types of metrics, suggesting not only higher accuracy and structural fidelity but also greater perceptual similarity to the original images. Additionally, the computational efficiency is better with fewer iterations. However, it is noteworthy that while fewer iterations result in higher metric scores and efficiency, visual inspection of the outputs indicates that larger iteration counts can produce a more varied range of reconstructed images for the same input, suggesting a potential tradeoff between the diversity of output and quantitative performance metrics.

\begin{table}[tb!]
	\centering
	\caption{Average reconstruction performances illustrating the effect of the iteration count for $236$ test images with $\alpha = 3$ and no aggregation across $5$ Monte Carlo runs.}
	    \begin{adjustbox}{width=\columnwidth}
     
		\begin{tabu}{@{\extracolsep{4pt}}cccccccc@{}}
			\hline
InDI Total&\multicolumn{3}{c}{Perceptual}&\multicolumn{2}{c}{Distortion}&Average\\
\cline{2-4} \cline{5-6}
Iteration Count ($T$)&FID $\downarrow$&LPIPS $\downarrow$&CLIP-IQA $\uparrow$&PSNR $\uparrow$&SSIM $\uparrow$&Runtime (sec.) $\downarrow$\\		
\hline
4& 100.96&0.20&0.77&26.78&0.73&1.11\\
8& 103.65&0.21&0.77&26.51&0.71&1.28\\
32& 109.36 & 0.22 & 0.77 & 26.08 & 0.68 & 2.41\\

			\hline					
	\end{tabu}
     \end{adjustbox}

\label{table:indinononoresultsiterationcount}
 
\end{table}

\subsection{Effect of Aggregation}


The impact of aggregation on reconstruction quality is significant, as evidenced by improvements in both perceptual and distortion metrics with an increasing number of combined reconstructions.
Table \ref{table:indinononoresultsensemblingeffect} demonstrates the significance of aggregation, with improvements observed in both perceptual and distortion metrics as the number of combined reconstructions increases. This simultaneous improvement across both types of metrics suggests that the aggregation approach does not conform to the typical constraints of the perception-distortion tradeoff space, where improvements in one metric are often countered by compromises in another. These results indicate that our method is not operating within a Pareto optimal region of this tradeoff space; that is, improvement is achieved in both perceptual and distortion metrics without the expected tradeoff.

The success of aggregation in improving these metrics can be attributed to its ability to combine information from multiple reconstructions into a single output and effectively average out errors and artifacts specific to individual outputs. Hence, this process not only increases the overall fidelity and structural integrity of the final image but also preserves the reliable patches of each reconstruction while reducing the impact of any individual output’s weakness.

We chose to present results for $T=32$ because it yields the best reconstruction quality, as it was the highest timestep we trained our pipeline for. At this timestep, reconstructions become more diverse, enhancing the benefits of aggregation. The diversity at $T=32$ case aligns with the core principle of InDI, which emphasizes iterative improvement over a series of smaller steps rather than predicting a clean target image in a single step. In image reconstruction, an inherently ill-posed problem, multiple high-quality images can serve as plausible reconstructions of a given low-quality input. Consequently, the outcome of a single-step regression model often aggregates various potential explanations, resulting in outputs that may lack detail and realism. The primary advantage of InDI is its approach of gradually refining the image through many smaller steps, ultimately leading to improved perceptual quality.

\begin{table}[tb!]
	\centering
	\caption{Average reconstruction performances showing the effect of the aggregation for $236$ test images under $\alpha = 3$ and $T=32$ setting ($5$ Monte Carlo runs).}
	    \begin{adjustbox}{width=\columnwidth}
     
		\begin{tabu}{@{\extracolsep{4pt}}cccccccc@{}}
			\hline
Number of Different &\multicolumn{3}{c}{Perceptual}&\multicolumn{2}{c}{Distortion}&Average\\
\cline{2-4} \cline{5-6}
Reconstructions ($2p$)&FID $\downarrow$&LPIPS $\downarrow$&CLIP-IQA $\uparrow$&PSNR $\uparrow$&SSIM $\uparrow$&Runtime (sec.) $\downarrow$\\
\hline

no aggregation& 109.36 & 0.22 & 0.77 & 26.08 & 0.68 & 2.41\\
4& 96.54 & 0.17 & 0.76 & 27.59 & 0.77 & 10.24\\
6& 95.01 & 0.16 & 0.76 & 27.83 & 0.79 & 13.55\\
8& 94.26 & 0.16 & 0.76 & 27.95 & 0.80 & 20.73\\
12& 93.64 & 0.15 & 0.76 & 28.18 & 0.81 & 29.24\\
24& 93.12 & 0.15 & 0.76 & 28.37 & 0.82 & 56.51\\

			\hline					
	\end{tabu}
     \end{adjustbox}

\label{table:indinononoresultsensemblingeffect}
 
\end{table}

However, the computational demands increase with the number of combined reconstructions, which is particularly important in scenarios where speed is crucial. This indicates a tradeoff between improved image quality and processing efficiency.


\subsection{Uncertainty Quantification Through Ensemble Outputs}

\begin{figure}[b]
    \centering

       \includegraphics[width=\linewidth]{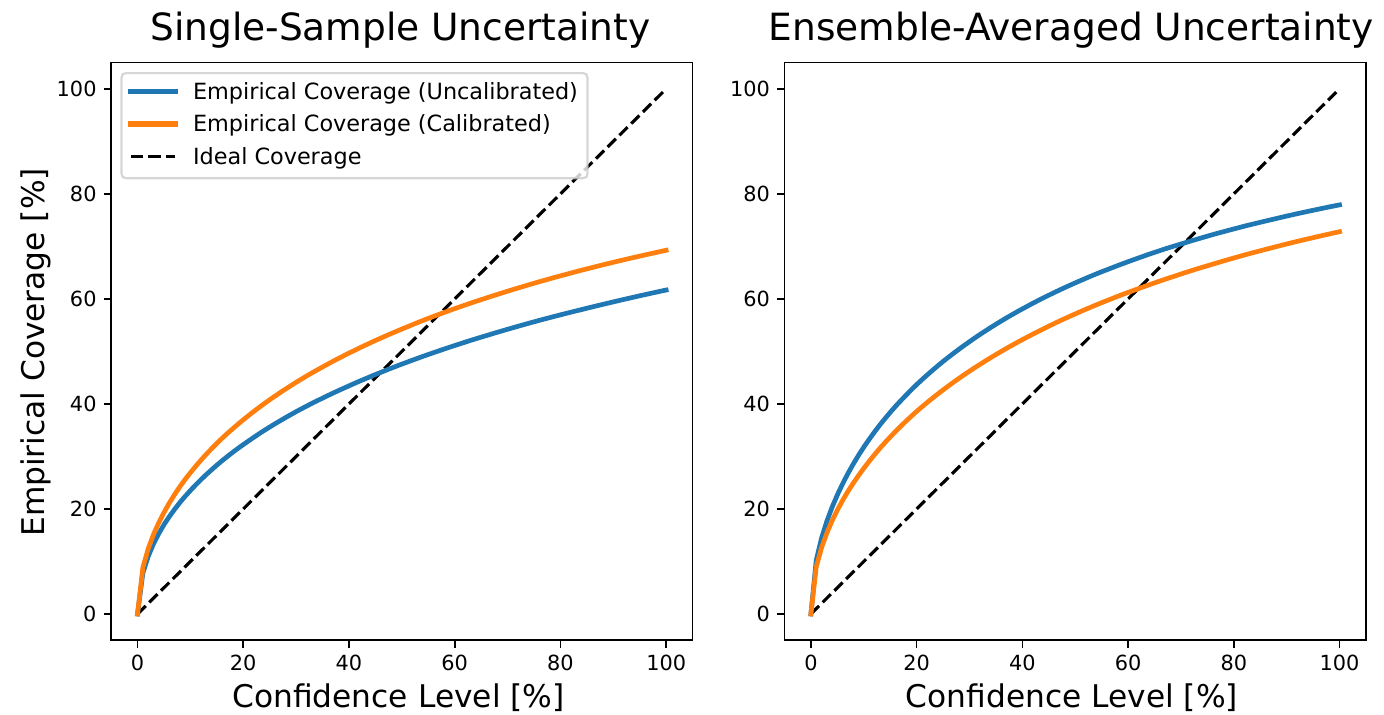}

    \caption{Calibration curves for two different cases: for a single output of the algorithm $\hat{\mathbf{x}}_{\text{final}}$ (left), the ensemble average of many output samples $\bar{\mathbf{x}}_{\text{final}}$ (right).}
    \label{fig:indinononocalibcurves}
\end{figure}

Uncertainty quantification is crucial for assessing the reliability of the outputs in image reconstruction, particularly in scenarios where decisions are based on the reconstructed images. Our stochastic pipeline inherently produces multiple reconstructions, enabling effective uncertainty quantification through ensemble analysis.
We estimate the uncertainty in our final reconstruction by calculating the variance for the ensemble of generated outputs. 
This variance approximates the expected squared error between the true image \(\mathbf{x}\) and the final reconstructed mean image \(\bar{\mathbf{x}}_{\text{final}}\), particularly since we anticipate that \(\mathbf{x} \approx \bar{\mathbf{x}}_{\text{final}}\) in our high-quality reconstructions. Here, \(\bar{\mathbf{x}}_{\text{final}}\) represents the mean of the ensemble of reconstructed images, which can be considered a random variable denoted as \(\hat{\mathbf{x}}_{\text{final}}\). The empirical variance serves as a Monte Carlo estimator for the expectation. The relationship is illustrated in the formula:
\begin{equation}
\mathbb{E}\{ \| \mathbf{x} - \hat{\mathbf{x}}_{\text{final}} \|_2^2 \} 
\approx 
\mathbb{E}\{ \| \bar{\mathbf{x}}_{\text{final}} - \hat{\mathbf{x}}_{\text{final}} \|_2^2 \} 
\approx 
\text{Var} \{  \hat{\mathbf{x}}_{\text{final}}^{(q)} \}^{2p}_{q=1}.
\end{equation}


This statistical approach leverages the diversity within the ensemble to predict uncertainty and capture variations that might not be evident from a single output.
This variance can be also used to provide an estimate of the error for the final result obtained from the ensemble average.

\begin{figure}[b]
    \centering

    \subfloat[]{%
        \begin{minipage}[t]{0.3333\columnwidth}
            \centering
            Actual\\error\\($\log$-scaled)\\[2pt]
            \includegraphics[width=\linewidth]{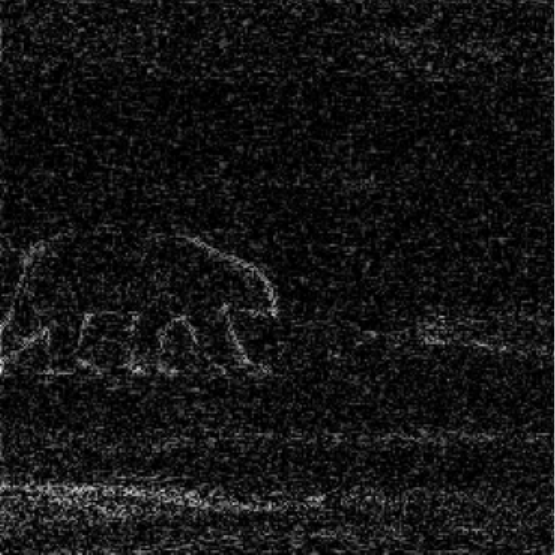}
        \end{minipage}
    }
    \subfloat[]{%
        \begin{minipage}[t]{0.3333\columnwidth}
            \centering
            Predicted uncertainty\\
            (before calibration)\\[2pt]
            \includegraphics[width=\linewidth]{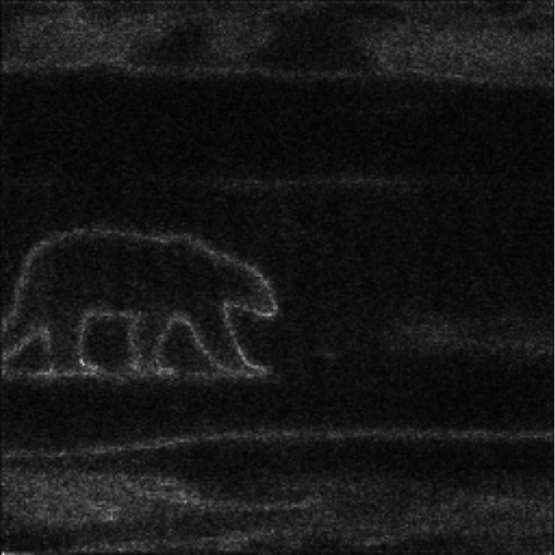}
        \end{minipage}
    }
    \subfloat[]{%
        \begin{minipage}[t]{0.3333\columnwidth}
            \centering
            Predicted uncertainty\\
            (after calibration)\\[2pt]
            \includegraphics[width=\linewidth]{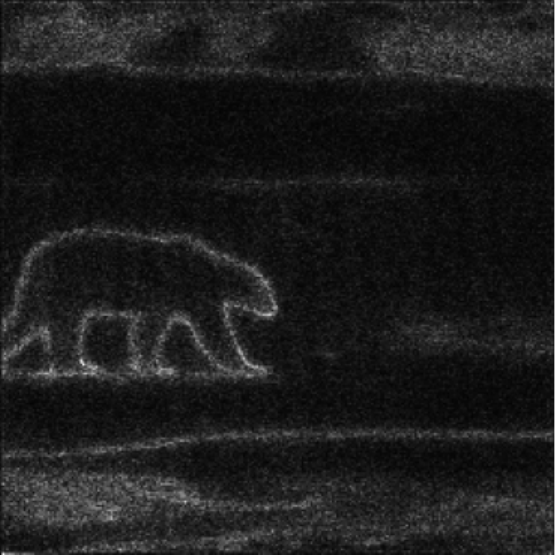}
        \end{minipage}
    }

    \vspace{-15pt}

    \subfloat[]{%
        \includegraphics[width=0.3333\columnwidth]{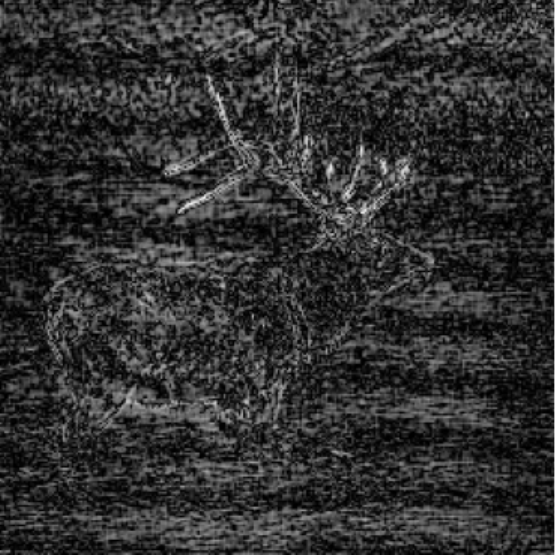}
    }
    \subfloat[]{%
        \includegraphics[width=0.3333\columnwidth]{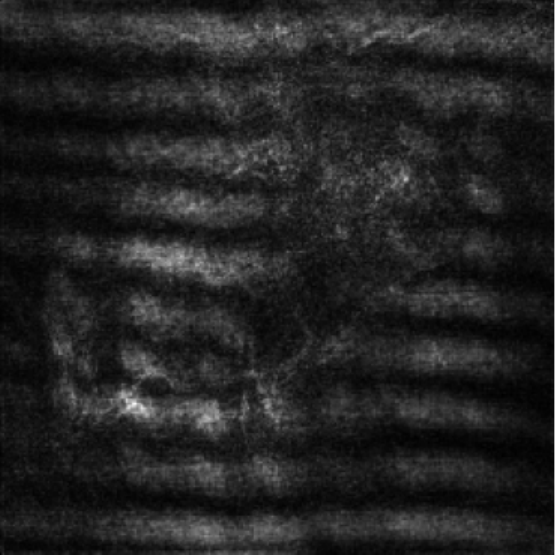}
    }
    \subfloat[]{%
        \includegraphics[width=0.3333\columnwidth]{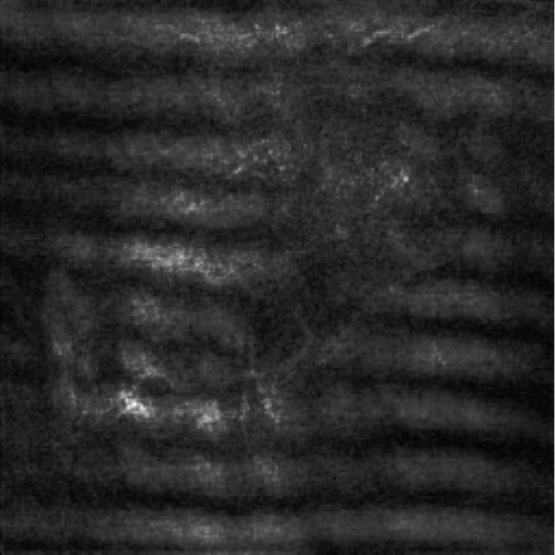}
    }

    \caption{Example uncertainty predictions and actual errors for the ensemble average of many output samples.}
    \label{fig:indinononocalibrationexample}
\end{figure}

While the empirical variance provides a useful estimate of uncertainty, it may not always perform optimally in all scenarios. Therefore, to ensure that the confidence levels assigned by our model accurately reflect its performance, calibration of the probabilistic predictions becomes essential. To achieve this, we utilize isotonic regression, a technique that excels in addressing monotonic distortions in data. Isotonic regression works by fitting a nondecreasing function to the predicted probabilities, allowing us to better align these probabilities with the actual outcomes \cite{NiculescuMizil2005PredictingGP}. This process enhances the calibration accuracy, which is particularly important in applications where decisions are based on the model's predictions. In our approach, we train the isotonic regression model using the predicted uncertainties derived from the empirical variance as input, with the actual errors associated with the reconstructed images serving as the target output using our validation dataset.
By regressing the actual errors against the predicted uncertainties, the isotonic regression model effectively learns to adjust the predicted confidence levels, enhancing overall calibration accuracy. Without proper calibration, uncalibrated predictions can lead to misleading conclusions and poor decision-making, as they may not truly represent the uncertainty associated with the reconstructed images.


Figure \ref{fig:indinononocalibcurves} presents the calibration curves for both uncalibrated and calibrated cases when using \(2p=24\) ensembles. Calibration curves are graphical representations that show how well the predicted uncertainty levels match the actual error rates encountered in our reconstructions. Empirical coverage refers to the proportion of instances where the true errors fall within the predicted uncertainty bounds. For instance, if our model predicts a certain confidence level for an output, we would expect that, at that confidence level, a corresponding percentage of actual errors should fall within that predicted range. Ideally, if the model is perfectly calibrated, the curve should align with a diagonal line indicating that the predicted probabilities match the observed outcomes.


The ideal coverage line in the figure serves as a benchmark for this perfect scenario. In our results, the calibrated curve shows improved empirical coverage compared to the uncalibrated one, indicating that the model's confidence levels now more accurately reflect the likelihood of error in its predictions. Furthermore, Fig. \ref{fig:indinononocalibrationexample} provides a visual comparison of actual reconstruction errors and the corresponding predicted uncertainties before and after calibration, demonstrating the effectiveness of the isotonic regression technique in improving our model's uncertainty estimation performance.

\subsection{Ablation Study: Acceleration Mechanism}

\begin{figure}[h]
    \centering
    \includegraphics[width=\linewidth]{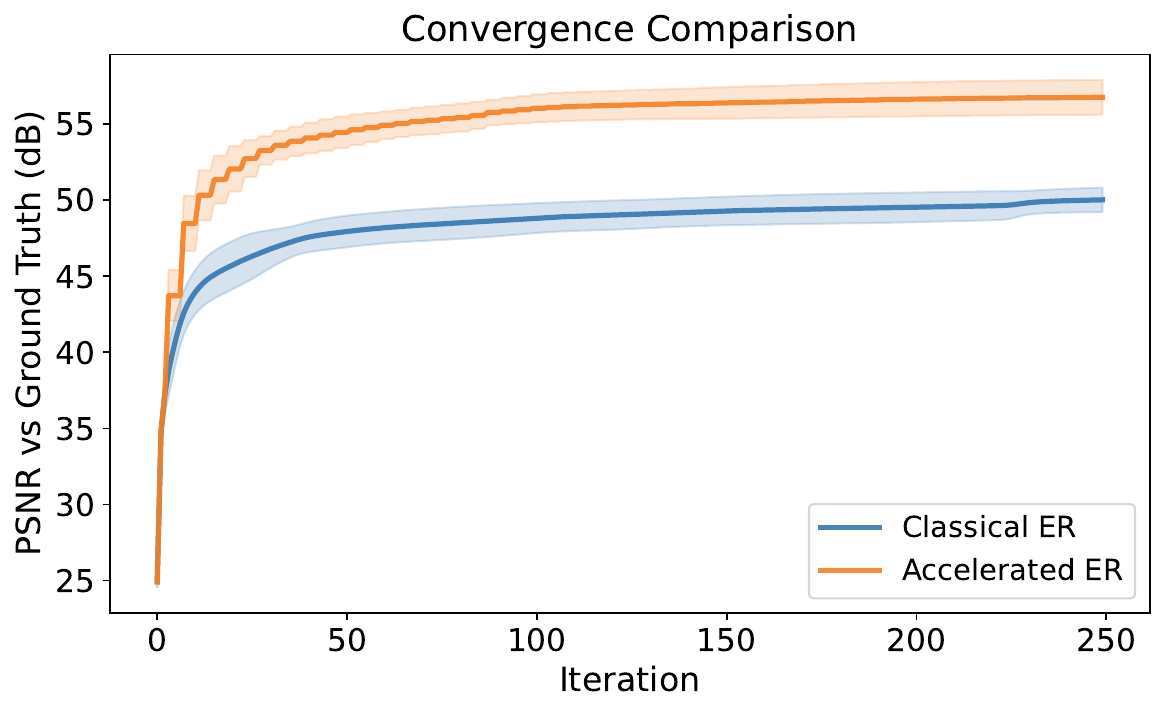}
    \caption{
    Convergence comparison of classical ER vs. Accelerated ER under noiseless measurements.
    We evaluate reconstruction quality (PSNR) against the ground truth image across 10 random initializations over 250 iterations. 
    Solid lines represent mean PSNR, while shaded regions indicate standard deviation across trials. 
    Accelerated ER achieves $56.8 \pm 1.2$ dB compared to $50.1 \pm 0.9$ dB for classical ER, representing a $+6.7$ dB improvement and demonstrating faster convergence.
    }
    \label{fig:AccelerationMechanism}
\end{figure}

To isolate the contribution of our proposed acceleration mechanism, we conduct a controlled ablation experiment comparing classical Error Reduction (ER) with Accelerated ER (AER) under noiseless measurements. We evaluate convergence behavior across 10 random initializations over 250 iterations. Figure~\ref{fig:AccelerationMechanism} demonstrates that Accelerated ER achieves substantially faster convergence and higher final reconstruction quality.

AER achieves a final PSNR of $56.8 \pm 1.2$ dB compared to $50.1 \pm 0.9$ dB for classical ER, representing a $+6.7$ dB improvement. The shaded regions indicate variance across trials, with both methods showing consistent convergence behavior.
This validates that the acceleration mechanism provides meaningful gains during the initialization stage, motivating its integration into our full iterative refinement pipeline.


\section{Conclusion}
\label{sec:conclusionindi}

In this paper, we have presented a novel approach to phase retrieval by integrating the Inversion by Direct Iteration (InDI) framework with advanced initialization and aggregation techniques. The learned image-to-image  pipeline improves the reconstruction through successive denoising
and data consistency stages. This method offers significant improvements over current approaches that typically start from random noise, resulting in enhanced reconstruction quality and computational efficiency. By leveraging initial crude estimates, our approach accelerates the training process, focusing computational resources on refining image reconstructions.

The use of aggregation in this context has proven particularly beneficial in achieving superior perceptual and distortion metrics, demonstrating that our method can advance image quality without the perception-distortion tradeoff. This opens up opportunities for future work to further advance this framework and push towards the Pareto optimal frontier.

Moreover, the extensive evaluations presented in this work highlight the robustness of our approach under various noise levels, outperforming established methods in terms of both accuracy and efficiency. As a result, this work sets a new benchmark for phase retrieval performance.


In conclusion, this paper demonstrates that integrating robust initialization strategies, geometric self-ensemble, aggregation techniques, and measurement consistency within the InDI framework leads to more efficient and accurate phase retrieval. The proposed approach can benefit a broad range of scientific and industrial applications. Future work may explore further improvements by more tightly integrating I2I denoising with model-based inference as well as extensions to more general nonlinear inverse problems.

\printcredits

\section*{Acknowledgments}
This work was supported in part by Scientific and Technological Research Council of Turkey (TUBITAK) under the Grant Number 120E505 and
by the BAGEP Award of the Science Academy.
Figen S. Oktem thanks TUBITAK and the Science Academy for the support.

\bibliographystyle{cas-model2-names}

\bibliography{bibliography}

@InProceedings{Mirza2023LearningFD,
  author = {Muhammad Usama Mirza and Onat Dalmaz and Hasan Atakan Bedel and Gokberk Elmas and Yilmaz Korkmaz and Alper Gungor and Salman UH Dar and Tolga Cukur},
  title  = {Learning Fourier-Constrained Diffusion Bridges for MRI Reconstruction},
  year   = {2023},
}

@Article{delbracio2023inversion,
  author  = {Delbracio, Mauricio and Milanfar, Peyman},
  title   = {Inversion by direct iteration: An alternative to denoising diffusion for image restoration},
  volume  = {2023},
  journal = {Trans. Mach. Learn. Res.},
  year    = {2023},
}

@InProceedings{kaya2025_eusipco,
  author    = {Kaya, Mehmet Onurcan and Oktem, Figen S},
  booktitle = {Proceedings of the 33rd European Signal Processing Conference},
  title     = {{I2I-PR}: Data-Driven Phase Retrieval Using Image-to-Image Diffusion Models},
  pages     = {1697--1701},
  year      = {2025},
}

@Article{personaFienup:13,
  author    = {James R. Fienup},
  title     = {Phase retrieval algorithms: a personal tour},
  number    = {1},
  pages     = {45--56},
  volume    = {52},
  journal   = {Appl. Opt.},
  month     = {1},
  IGNOREpublisher = {Optica Publishing Group},
  year      = {2013},
}

@Article{Isil:19,
  author    = {Cagatay Isil and Figen S. Oktem and Aykut Koç},
  title     = {Deep iterative reconstruction for phase retrieval},
  number    = {20},
  pages     = {5422--5431},
  volume    = {58},
  journal   = {Appl. Opt.},
  month     = {7},
  IGNOREpublisher = {Optica Publishing Group},
  year      = {2019},
}

@Article{shechtman2015phase,
  author    = {Shechtman, Yoav and Eldar, Yonina C and Cohen, Oren and Chapman, Henry Nicholas and Miao, Jianwei and Segev, Mordechai},
  title     = {Phase retrieval with application to optical imaging: a contemporary overview},
  number    = {3},
  pages     = {87--109},
  volume    = {32},
  journal   = {IEEE Signal Processing Magazine},
  IGNOREpublisher = {IEEE},
  year      = {2015},
}

@Article{heckel2024deep,
  title={Deep learning for accelerated and robust MRI reconstruction},
  author={Heckel, Reinhard and Jacob, Mathews and Chaudhari, Akshay and Perlman, Or and Shimron, Efrat},
  journal={Magnetic Resonance Materials in Physics, Biology and Medicine},
  volume={37},
  number={3},
  pages={335--368},
  year={2024},
  IGNOREIGNOREpublisher={Springer}
}

@Article{Whang2021DeblurringVS,
  author  = {Jay Whang and Mauricio Delbracio and Hossein Talebi and Chitwan Saharia and Alexandros G. Dimakis and Peyman Milanfar},
  title   = {Deblurring via Stochastic Refinement},
  pages   = {16272-16282},
  journal = {2022 IEEE/CVF Conference on Computer Vision and Pattern Recognition},
  year    = {2021},
}

@Article{Wang_2024,
  author    = {Wang, Kaiqiang and Song, Li and Wang, Chutian and Ren, Zhenbo and Zhao, Guangyuan and Dou, Jiazhen and Di, Jianglei and Barbastathis, George and Zhou, Renjie and Zhao, Jianlin and Lam, Edmund Y.},
  title     = {On the use of deep learning for phase recovery},
  issn      = {2047-7538},
  number    = {1},
  volume    = {13},
  journal   = {Light: Science \& Applications},
  month     = jan,
  IGNOREpublisher = {Springer Science and Business Media LLC},
  year      = {2024},
}

@Article{Maiden:17,
  author    = {Andrew Maiden and Daniel Johnson and Peng Li},
  title     = {Further improvements to the ptychographical iterative engine},
  number    = {7},
  pages     = {736--745},
  volume    = {4},
  journal   = {Optica},
  month     = {7},
  IGNOREpublisher = {Optica Publishing Group},
  year      = {2017},
}

@InProceedings{Loshchilov2017DecoupledWD,
  author    = {Ilya Loshchilov and Frank Hutter},
  booktitle = {International Conference on Learning Representations},
  title     = {Decoupled Weight Decay Regularization},
  year      = {2017},
}

@Article{Blau2017ThePT,
  author  = {Yochai Blau and Tomer Michaeli},
  title   = {The Perception-Distortion Tradeoff},
  pages   = {6228-6237},
  journal = {2018 IEEE/CVF Conference on Computer Vision and Pattern Recognition},
  year    = {2017},
}

@Article{Uelwer2019PhaseRU,
  author  = {Tobias Uelwer and Alexander Oberstrass and Stefan Harmeling},
  title   = {Phase Retrieval Using Conditional Generative Adversarial Networks},
  pages   = {731-738},
  journal = {2020 25th International Conference on Pattern Recognition (ICPR)},
  year    = {2019},
}

@Article{Aggarwal2017MoDLMD,
  author  = {Hemant Kumar Aggarwal and Merry P. Mani and Mathews Jacob},
  title   = {MoDL: Model-Based Deep Learning Architecture for Inverse Problems},
  pages   = {394-405},
  volume  = {38},
  journal = {IEEE Transactions on Medical Imaging},
  year    = {2017},
}

@InProceedings{martin2001bsd,
  author       = {Martin, David and Fowlkes, Charless and Tal, Doron and Malik, Jitendra},
  booktitle    = {Proceedings of the IEEE International Conference on Computer Vision (ICCV)},
  title        = {A database of human segmented natural images and its application to evaluating segmentation algorithms and measuring ecological statistics},
  organization = {IEEE},
  pages        = {416--423},
  volume       = {2},
  year         = {2001},
}

@Article{Bansal2022ColdDI,
  author  = {Arpit Bansal and Eitan Borgnia and Hong-Min Chu and Jie Li and Hamid Kazemi and Furong Huang and Micah Goldblum and Jonas Geiping and Tom Goldstein},
  title={Cold diffusion: Inverting arbitrary image transforms without noise},
  journal={Advances in Neural Information Processing Systems},
  volume={36},
  pages={41259--41282},
  year={2023}
}

@Article{isil2024deep,
  author  = {Cagatay Isil and Figen S. Oktem},
journal = {Appl. Opt.},
number = {5},
pages = {A84--A94},
IGNOREIGNOREpublisher = {Optica Publishing Group},
title = {Deep plug-and-play HIO approach for phase retrieval},
volume = {64},
month = {Feb},
year = {2025},
}

@misc{diffusers,
  author      = {Patrick von Platen and Suraj Patil and Anton Lozhkov and Pedro Cuenca and Nathan Lambert and Kashif Rasul and Mishig Davaadorj and Dhruv Nair and Sayak Paul and Steven Liu and William Berman and Yiyi Xu and Thomas Wolf},
  title       = {Diffusers: State-of-the-art diffusion models},
  url         = {https://github.com/huggingface/diffusers},
  version     = {0.12.1},
  cff-version = {1.2.0},
  license     = {Apache-2.0},
  year        = {2024},
}

@Article{Fannjiang2020TheNO,
  author  = {Albert Fannjiang and Thomas Strohmer},
  title   = {The numerics of phase retrieval},
  pages   = {125 - 228},
  volume  = {29},
  journal = {Acta Numerica},
  year    = {2020},
}

@Article{AragnArtacho2019TheDA,
  author  = {Francisco J. Arag{\'o}n Artacho and Rub{\'e}n Campoy and Matthew K. Tam},
  title   = {The Douglas–Rachford algorithm for convex and nonconvex feasibility problems},
  pages   = {201 - 240},
  volume  = {91},
  journal = {Mathematical Methods of Operations Research},
  year    = {2019},
}

@Article{chung2024direct,
  author  = {Chung, Hyungjin and Kim, Jeongsol and Ye, Jong Chul},
  title   = {Direct diffusion bridge using data consistency for inverse problems},
  volume  = {36},
  journal = {Advances in Neural Information Processing Systems},
  year    = {2024},
}

@Article{Goy2018LowPC,
  author  = {Alexandre Goy and Kwabena Arthur and Shuai Li and George Barbastathis},
  title   = {Low Photon Count Phase Retrieval Using Deep Learning.},
  pages   = {243902},
  volume  = {121 24},
  journal = {Physical review letters},
  year    = {2018},
}

@Article{Loshchilov2016SGDRSG,
title={{SGDR}: Stochastic Gradient Descent with Warm Restarts},
author={Ilya Loshchilov and Frank Hutter},
booktitle={International Conference on Learning Representations},
year={2017},
}

@InProceedings{Isil:20,
  author    = {Cagatay Isil and Figen S. Oktem},
  booktitle = {Imaging and Applied Optics Congress},
  title     = {Model-based Phase Retrieval with Deep Denoiser Prior},
  IGNOREpublisher = {Optica Publishing Group},
  journal   = {Imaging and Applied Optics Congress},
  year      = {2020},
  pages={CF2C--5},
}

@Article{wang2004image,
  author    = {Wang, Zhou and Bovik, Alan C and Sheikh, Hamid R and Simoncelli, Eero P},
  title     = {Image quality assessment: from error visibility to structural similarity},
  number    = {4},
  pages     = {600--612},
  volume    = {13},
  journal   = {IEEE Transactions on Image Processing},
  IGNOREpublisher = {IEEE},
  year      = {2004},
}

@Article{marcheshrinkwrap,
  author  = {Marchesini, Stefano and He, H. and Chapman, Henry and Hau-Riege, S.P. and Noy, Aleksandr and Howells, Malcolm and Weierstall, Uwe and Spence, J},
  title   = {X-ray image reconstruction from a diffraction pattern alone},
  volume  = {68},
  journal = {Physical Review B},
  month   = {07},
  year    = {2003},
}

@InProceedings{kawar2022denoising,
 author = {Kawar, Bahjat and Elad, Michael and Ermon, Stefano and Song, Jiaming},
 booktitle = {Advances in Neural Information Processing Systems},
 pages = {23593--23606},
 title = {Denoising Diffusion Restoration Models},
 volume = {35},
 year = {2022}
}

@InProceedings{Liu2023I2SBIS,
  author    = {Guan-Horng Liu and Arash Vahdat and De-An Huang and Evangelos A. Theodorou and Weili Nie and Anima Anandkumar},
  booktitle = {International Conference on Machine Learning},
  title     = {I2SB: Image-to-Image Schr{\"o}dinger Bridge},
  year      = {2023},
articleno = {915},
numpages = {21},

}

@Article{lpips_Zhang2018TheUE,
  author  = {Richard Zhang and Phillip Isola and Alexei A. Efros and Eli Shechtman and Oliver Wang},
  title   = {The Unreasonable Effectiveness of Deep Features as a Perceptual Metric},
  pages   = {586-595},
  journal = {2018 IEEE/CVF Conference on Computer Vision and Pattern Recognition},
  year    = {2018},
}

@Article{ma2017waterloo,
  author  = {Ma, Kede and Duanmu, Zhengfang and Wu, Qingbo and Wang, Zhou and Yong, Hongwei and Li, Hongliang and Zhang, Lei},
  title   = {{Waterloo Exploration Database}: New Challenges for Image Quality Assessment Models},
  number  = {2},
  pages   = {1004--1016},
  volume  = {26},
  journal = {IEEE Transactions on Image Processing},
  month   = {2},
  year    = {2017},
}

@Article{franzen,
  author  = {Franzen, Richard W.},
  title   = {True Color Kodak Images},
  journal = {http://r0k.us/graphics/kodak},
}

@Article{hayes1982,
  author   = {M. Hayes},
  title    = {The reconstruction of a multidimensional sequence from the phase or magnitude of its Fourier transform},
  issn     = {0096-3518},
  number   = {2},
  pages    = {140-154},
  volume   = {30},
  journal  = {IEEE Transactions on Acoustics, Speech, and Signal Processing},
  month    = {4},
  year     = {1982},
}

@Article{1963_walther_pr_optics,
  author    = {Adriaan Walther},
  title     = {The Question of Phase Retrieval in Optics},
  number    = {1},
  pages     = {41-49},
  volume    = {10},
  journal   = {Optica Acta: International Journal of Optics},
  IGNOREpublisher = {Taylor & Francis},
  year      = {1963},
}

@InProceedings{deng2009imagenet,
  author       = {Deng, Jia and Dong, Wei and Socher, Richard and Li, Li-Jia and Li, Kai and Fei-Fei, Li},
  booktitle    = {Proceedings of the IEEE Computer Vision and Pattern Recognition},
  title        = {Imagenet: A large-scale hierarchical image database},
  organization = {IEEE},
  pages        = {248--255},
  year         = {2009},
}

@Article{lopeztaipa,
  author  = {López-Tapia, Santiago and Molina, Rafael and Katsaggelos, Aggelos},
  title   = {Deep learning approaches to inverse problems in imaging: Past, present and future},
  pages   = {103285},
  volume  = {119},
  journal = {Digital Signal Processing},
  month   = {10},
  year    = {2021},
}

@Article{fienup1978reconstruction,
  author    = {Fienup, James R},
  title     = {Reconstruction of an object from the modulus of its Fourier transform},
  number    = {1},
  pages     = {27--29},
  volume    = {3},
  journal   = {Optics letters},
  IGNOREpublisher = {Optical Society of America},
  year      = {1978},
}

@inproceedings{fid_Heusel2017GANsTB,
  author  = {Martin Heusel and Hubert Ramsauer and Thomas Unterthiner and Bernhard Nessler and G{\"u}nter Klambauer and Sepp Hochreiter},
  title   = {GANs Trained by a Two Time-Scale Update Rule Converge to a Nash Equilibrium},
year = {2017},
IGNOREisbn = {9781510860964},
IGNOREIGNOREpublisher = {Curran Associates Inc.},
IGNOREaddress = {Red Hook, NY, USA},
IGNOREabstract = {Generative Adversarial Networks (GANs) excel at creating realistic images with complex models for which maximum likelihood is infeasible. However, the convergence of GAN training has still not been proved. We propose a two time-scale update rule (TTUR) for training GANs with stochastic gradient descent on arbitrary GAN loss functions. TTUR has an individual learning rate for both the discriminator and the generator. Using the theory of stochastic approximation, we prove that the TTUR converges under mild assumptions to a stationary local Nash equilibrium. The convergence carries over to the popular Adam optimization, for which we prove that it follows the dynamics of a heavy ball with friction and thus prefers flat minima in the objective landscape. For the evaluation of the performance of GANs at image generation, we introduce the 'Fr\'{e}chet Inception Distance" (FID) which captures the similarity of generated images to real ones better than the Inception Score. In experiments, TTUR improves learning for DCGANs and Improved Wasserstein GANs (WGAN-GP) outperforming conventional GAN training on CelebA, CIFAR-10, SVHN, LSUN Bedrooms, and the One Billion Word Benchmark.},
booktitle = {Proceedings of the 31st International Conference on Neural Information Processing Systems},
pages = {6629–6640},
numpages = {12},
IGNORElocation = {Long Beach, California, USA},
IGNOREseries = {NIPS'17}
}

@Article{marchesini2007invited,
  author    = {Marchesini, Stefano},
  title     = {Invited article: A unified evaluation of iterative projection algorithms for phase retrieval},
  number    = {1},
  volume    = {78},
  journal   = {Review of scientific instruments},
  IGNOREpublisher = {AIP Publishing},
  year      = {2007},
}

@Article{fienup1982comparison,
  author    = {J. R. Fienup},
  title     = {Phase retrieval algorithms: a comparison},
  number    = {15},
  pages     = {2758--2769},
  volume    = {21},
  journal   = {Appl. Opt.},
  month     = {8},
  IGNOREpublisher = {OSA},
  year      = {1982},
}

@Article{NiculescuMizil2005PredictingGP,
  author  = {Alexandru Niculescu-Mizil and Rich Caruana},
  title   = {Predicting good probabilities with supervised learning},
  journal = {Proceedings of the 22nd international conference on Machine learning},
  year    = {2005},
}

@InProceedings{pmlr-v80-metzler18a,
  author    = {Metzler, Christopher and Schniter, Phillip and Veeraraghavan, Ashok and Baraniuk, Richard},
  booktitle = {Proceedings of the 35th International Conference on Machine Learning},
  title     = {pr{D}eep: Robust Phase Retrieval with a Flexible Deep Network},
  pages     = {3498--3507},
  IGNOREpublisher = {JMLR.org},
  year      = {2018},
}

@Article{gs1978,
  author  = {Gerchberg, R. W. and Saxton, W. O.},
  title   = {A practical algorithm for the determination of phase from image and diffraction plane pictures},
  pages   = {237-250},
  volume  = {35},
  journal = {Optik},
  month   = {11},
  year    = {1972},
}

@Article{stefanoqianpty,
  author  = {Qian, Jianliang and Yang, C. and Schirotzek, A. and Maia, Filipe and Marchesini, Stefano},
  title   = {Efficient algorithms for ptychographic phase retrieval, in Inverse Problems and Applications},
  pages   = {261-280},
  volume  = {615},
  journal = {Contemp. Math},
  month   = {01},
  year    = {2014},
}

@InProceedings{zhang2017learning,
  author    = {Zhang, Kai and Zuo, Wangmeng and Gu, Shuhang and Zhang, Lei},
  booktitle = {Proceedings of the IEEE Conference on Computer Vision and Pattern Recognition},
  title     = {Learning Deep {CNN} Denoiser Prior for Image Restoration},
  pages     = {3929--3938},
  IGNOREpublisher = {IEEE},
  year      = {2017},
}

@Article{Dong_2023,
  author    = {Dong, Jonathan and Valzania, Lorenzo and Maillard, Antoine and Pham, Thanh-an and Gigan, Sylvain and Unser, Michael},
  title     = {Phase Retrieval: From Computational Imaging to Machine Learning: A tutorial},
  issn      = {1558-0792},
  number    = {1},
  pages     = {45–57},
  volume    = {40},
  journal   = {IEEE Signal Processing Magazine},
  month     = jan,
  IGNOREpublisher = {Institute of Electrical and Electronics Engineers (IEEE)},
  year      = {2023},
}

@InProceedings{clipiqa_Wang2022ExploringCF,
  author    = {Jianyi Wang and Kelvin C. K. Chan and Chen Change Loy},
  booktitle = {AAAI Conference on Artificial Intelligence},
  title     = {Exploring CLIP for Assessing the Look and Feel of Images},
  year      = {2022},
}

@Article{Saharia2021PaletteID,
  author  = {Chitwan Saharia and William Chan and Huiwen Chang and Chris A. Lee and Jonathan Ho and Tim Salimans and David J. Fleet and Mohammad Norouzi},
  title   = {Palette: Image-to-Image Diffusion Models},
  journal = {ACM SIGGRAPH 2022 Conference Proceedings},
  year    = {2021},
}

@inproceedings{liu2023flow,
title={Flow Straight and Fast: Learning to Generate and Transfer Data with Rectified Flow},
author={Xingchao Liu and Chengyue Gong and qiang liu},
booktitle={The Eleventh International Conference on Learning Representations},
year={2023},
}

@inproceedings{heitz2023iterative,
  title={Iterative $\alpha$-(de) blending: A minimalist deterministic diffusion model},
  author={Heitz, Eric and Belcour, Laurent and Chambon, Thomas},
  booktitle={ACM SIGGRAPH 2023 Conference Proceedings},
  year={2023}
}

@article{oppenheim2005importance,
  title={The importance of phase in signals},
  author={Oppenheim, Alan V and Lim, Jae S},
  journal={Proceedings of the IEEE},
  volume={69},
  number={5},
  pages={529--541},
  year={2005},
  IGNOREpublisher={IEEE}
}

@inproceedings{peer2022beyond,
  title={Beyond Griffin-Lim: Improved iterative phase retrieval for speech},
  author={Peer, Tal and Welker, Simon and Gerkmann, Timo},
  booktitle={2022 International Workshop on Acoustic Signal Enhancement (IWAENC)},
  pages={1--5},
  year={2022},
  IGNOREorganization={IEEE}
}

@article{sarao2020optimization,
  title={Optimization and generalization of shallow neural networks with quadratic activation functions},
  author={Sarao Mannelli, Stefano and Vanden-Eijnden, Eric and Zdeborov{\'a}, Lenka},
  journal={Advances in Neural Information Processing Systems},
  volume={33},
  pages={13445--13455},
  year={2020}
}

@article{latychevskaia2019iterative,
  title={Iterative phase retrieval for digital holography: tutorial},
  author={Latychevskaia, Tatiana},
  journal={Journal of the Optical Society of America A},
  volume={36},
  number={12},
  pages={D31--D40},
  year={2019},
  IGNOREpublisher={OSA}
}

@article{chang2018hybrid,
  title={Hybrid optical-electronic convolutional neural networks with optimized diffractive optics for image classification},
  author={Chang, Julie and Sitzmann, Vincent and Dun, Xiong and Heidrich, Wolfgang and Wetzstein, Gordon},
  journal={Scientific reports},
  volume={8},
  number={1},
  pages={12324},
  year={2018},
  IGNOREpublisher={Nature Publishing Group UK London}
}

@article{millane1990phase,
  title={Phase retrieval in crystallography and optics},
  author={Millane, Rick P},
  journal={Journal of the Optical Society of America A},
  volume={7},
  number={3},
  pages={394--411},
  year={1990},
  IGNOREpublisher={Optical Society of America}
}

@article{zheng2013wide,
  title={Wide-field, high-resolution Fourier ptychographic microscopy},
  author={Zheng, Guoan and Horstmeyer, Roarke and Yang, Changhuei},
  journal={Nature photonics},
  volume={7},
  number={9},
  pages={739--745},
  year={2013},
  IGNOREpublisher={Nature Publishing Group UK London}
}

@article{hubbleFienup:93,
author = {J. R. Fienup and J. C. Marron and T. J. Schulz and J. H. Seldin},
journal = {Appl. Opt.},
number = {10},
pages = {1747--1767},
publisher = {Optica Publishing Group},
title = {Hubble Space Telescope characterized by using phase-retrieval algorithms},
volume = {32},
year = {1993},
}

@article{ddrmprKaya:25,
author = {Mehmet Onurcan Kaya and Figen S. Oktem},
journal = {Appl. Opt.},
number = {5},
pages = {A95--A105},
title = {DDRM-PR: Fourier phase retrieval using denoising diffusion restoration models},
volume = {64},
year = {2025},
}

@INPROCEEDINGS{10477083,
  author={Shoushtari, Shirin and Liu, Jiaming and Kamilov, Ulugbek S.},
  booktitle={2023 57th Asilomar Conference on Signals, Systems, and Computers}, 
  title={Diffusion Models for Phase Retrieval in Computational Imaging}, 
  year={2023},
  volume={},
  number={},
  pages={779-783},
  }

@article{zhang2021plug,
  title={Plug-and-play image restoration with deep denoiser prior},
  author={Zhang, Kai and Li, Yawei and Zuo, Wangmeng and Zhang, Lei and Van Gool, Luc and Timofte, Radu},
  journal={IEEE Transactions on Pattern Analysis and Machine Intelligence},
  volume={44},
  number={10},
  pages={6360--6376},
  year={2021}
}


\bio{}
\textbf{Mehmet Onurcan Kaya} received the B.S. and M.S. degrees in electrical engineering from Middle East Technical University, Ankara, Turkey, in 2021 and 2024, respectively. He is currently pursuing a Ph.D. degree at the Technical University of Denmark, Department of Applied Mathematics and Computer Science. His research interests include machine learning, multimodal generative AI, computer vision, and computational imaging.
\endbio

\bio{}
\textbf{Figen S. Oktem} received the B.S. and M.S. degrees in electrical engineering from Bilkent University, Ankara, Turkey, in 2007 and 2009, respectively, and the Ph.D. degree in electrical and computer engineering from the University of Illinois at Urbana-Champaign (UIUC), Champaign, IL, USA, in 2014. She was then a Postdoctoral research associate with the NASA Goddard Space Flight Center, where she worked on high-resolution spectral imaging. She is currently an associate professor with the Department of Electrical and Electronics Engineering, Middle East Technical University, Ankara. Her research interests include computational imaging, inverse problems, statistical signal processing, machine learning, compressed sensing, and optical information processing. She was the recipient of the 2025 BAGEP Young Scientist Award (Science Academy, Turkey) and the 2025 L’Oréal-UNESCO For Women in Science National Award. At UIUC, she was awarded the NASA Earth and Space Science Fellowship and the Professor Kung Chie Yeh Endowed Fellowship.
\endbio


\end{document}